\documentclass[12pt,a4paper,openright]{article}
\usepackage{graphicx}
\usepackage{color}
\usepackage[T1]{fontenc}
\usepackage[latin2]{inputenc}
\usepackage{amsmath}
\usepackage{amssymb}
\newtheorem{theorem}{Theorem}
\newtheorem{proposition}[theorem]{Proposition}
\newtheorem{defi}[theorem]{Definition}
\newtheorem{lemma}[theorem]{Lemma}
\newtheorem{cor}[theorem]{Corollary}
\newtheorem{remark}[theorem]{Remark}
\newcommand{\bea}{\begin{array}}
\newcommand{\eea}{\end{array}}
\newcommand{\be}{\begin{equation}}
\newcommand{\ee}{\end{equation}}
\newcommand{\ba}{\begin{eqnarray}}
\newcommand{\ea}{\end{eqnarray}}
\newcommand{\baw}{\begin{eqnarray*}}
\newcommand{\eaw}{\end{eqnarray*}}
\newcommand{\bt}{\begin{tabular}}
\newcommand{\et}{\end{tabular}}
\newcommand{\bc}{\begin{center}}
\newcommand{\ec}{\end{center}}
\newcommand{\ben}{\begin{enumerate}}
\newcommand{\een}{\end{enumerate}}
\newcommand{\bi}{\begin{itemize}}
\newcommand{\ei}{\end{itemize}}
\newcommand{\bmpage}{\begin{minipage}}
\newcommand{\empage}{\end{minipage}}
\newcommand{\bft}{\begin{footnotesize}}
\newcommand{\eft}{\end{footnotesize}}

\newcommand{\brak}{\langle}
\newcommand{\kket}{\rangle}

\def\vs{\vskip.5cm}

\def \dg{\dagger}
\def \df {{\sf d}}

\def \inf {{\rm inf}\,}

\def \sup {{\rm sup}\,}

\def \Tr{{\rm Tr} }

\def\brak{\langle}
\def\kket{\rangle}

\def\La {\Lambda}

\def\jgr{{\bf j} }
\def\kgr{{\bf k} }
\def\lgr{{\bf l} }
\def\mgr{{\bf m} }
\def\0gr{{\bf 0} }

\def\xgr{{\bf x} }
\def\ygr{{\bf y} }
\def\zgr{{\bf z} }

\def\Zerogr{\bf 0}

\def\Bcal{ {\cal B}}

\def\Hcal{ {\cal H} }

\def\Cdb{ { \mathbb C}}

\def\Ndb{ { \mathbb N}}
\def\Rdb{ { \mathbb R}}

\def\Zdb{ { \mathbb Z}}

\newcommand{\vsm}{\vspace*{-0.5ex}}

\newcommand{\Dm}{{\mbox{${\cal{D}}_m$}}}
\newcommand{\DM}[1]{{\mbox{${\mathrm D}_{#1}$}}}
\newcommand{\MM}{{\mbox{${\mathbf M}$}}}
\newcommand{\UU}{u}
\newcommand{\DD}{{\cal D}}
\newcommand{\NN}{{\mathrm N}}

\newcommand{\pG}{{\mathrm G}}
\newcommand{\pH}{{\mathrm H}}
\newcommand{\pT}{{\mathrm T}}
\newcommand{\pS}{{\mathrm S}}
\newcommand{\lrs}{N}
\newcommand{\dodgrup}{\dot{+}}
\newcommand{\odejgrup}{\dot{-}}
\newcommand{\dsp}{\displaystyle}
 
\begin{document}

\title{Non-existence of Bose-Einstein condensation in Bose-Hubbard model 
in dimensions 1 and 2}
\author{\em
Piotr Stachura${}^*$, 
Wies\l aw Pusz${}^\dg$,
Jacek Wojtkiewicz${}^\dg$,
  \\
  \\
   ${}^*$
   Institute of Information Technology,
   \\
Warsaw University of Life Sciences--SGGW,\\
ul. Nowoursynowska 159, 02-776 Warszawa, Poland\\
e-mail: ${\rm piotr\_{}stachura1@sggw.pl}$ 
 \\
\\
  ${}^\dg$ Department for Mathematical Methods in Physics,\\ Faculty of 
Physics, Warsaw University,\\
   Pasteura 5, 02-093 Warszawa, Poland\\
e-mail:  ${\rm wpusz@fuw.edu.pl}$ 
(W.P.), ${\rm wjacek@fuw.edu.pl}$ (J.W.)
  }
  \maketitle
\abstract{
We apply the Bogoliubov inequality to the Bose-Hubbard model to rule out 
the possibility of Bose-Einstein condensation.  The result
holds in one and two dimensions, for any filling at any
nonzero temperature. This result can be considered as complementary to
analogous, classical result known for interacting bosons in continuum.
}

\section{Introduction}
Many-body boson systems were among first quantum systems, where the problem of phase transitions has been noticed. In a system of non-interacting bosons 
it was investigated in the very beginning of quantum theory \cite{Bose}, \cite{Einstein1}, \cite{Einstein2}, \cite{Einstein3}. 
For an excellent review of non-interacting boson systems see \cite{ZUK}.
 
In systems of {\em interacting} bosons, the problem turned out much more difficult. Significant progress has been made in late 40's of XX-th century, 
when the existence of phase transition in 3d systems has been (non-rigorously) shown by Bogolyubov \cite{Bogolubow47}. 
Further contributions into development of the  theory have been made, among others, by Penrose, Onsager, \cite{Penrose}, \cite{PenroseOnsager},  
Feynman \cite{Feynman}, Lee, Huang, Yang \cite{Yang}, \cite{Huang}. (For exhaustive reviews  see \cite{ZagrebnovBru}, \cite{CDZ}).

The Bogolyubov approach was heuristic one and it cannot be regarded as a rigorous proof of the Bose-Einstein condensation. 
It has been based on profound understanding of physics of the problem, however, it lacked a rigorous justification, 
which has been made only half century later (but only in certain set of physical parameters) \cite{LSSY}. 
(For another approach, assuming however fulfilling of certain boundary conditions, see \cite{VerbrEtAl}).

The situation in lower dimensions, i.e. 1 and 2, turned out to be more tractable. In the sixties of XX-th century 
there appeared convincing explanation that in systems of continuous bosons there is no condensation at positive temperatures \cite{Hohenberg}. 
This paper was based on deep physical insight, but it also cannot be considered as a proof -- even the Hamiltonian didn't appear in the paper!
About ten years later, Bouziane and Martin in \cite{BouzianeMartin}  analysed the problem 
rigorously and their results can be considered as a 
proof of the lack of Bose-Einstein condensation in dimensions one and two.

Technical tool, being the cornerstone of the proof by  Bouziane and Martin, it was  (certain version of) {\em the Bogolyubov inequality}.
 
 It was used in a whole series of papers, where non-existence of Long Range Order (LRO) has been proven 
in systems possessing the continuous symmetry group. Boson systems also belong to this class. 
More precisely,  celebrated  Mermin-Wagner theorem \cite{MerminWagner} tells that in  lattice spin systems fulfilling the following assumptions: 
 \ben
\item they possess continuous symmetry group;
\item dimension of the lattice is  1 or 2; 
\item the interactions exhibit sufficiently fast spatial fall-off,
\een
there can be no LRO at positive temperatures.

This theorem -- half-century-old at present -- is one of few general beautiful and powerful results, 
concerning the (im)possibility of ordering  in lattice systems. It was generalized and extended in numerous directions; for some of these achievements, see \cite{Simon}. 

The Mermin-Wagner theorem has been proven with the use of Bogolyubov inequality \cite{Bogolubow62}. For (finite) spin or 
fermion systems this is a {\em matrix} inequality, whose proof is tricky but elementary.  For bosonic systems, however, 
one encounters technical complications due to the fact that the bosonic Hilbert space is infinite dimensional even for 
finite lattices. One is forced to develop suitable {\em operator} variant of Bogolyubov inequality. It was the case of the paper
 \cite{BouzianeMartin}, where it has been proven certain version of Bogolyubov inequality, whose application resulted in the proof of 
lack of Bose-Einstein condensation in continuous systems in dimensions one and two. Another operator versions of the Bogolyubov 
inequality have also been proven in papers \cite{PaChor} and \cite{WPS2}, where they have been applied to prove lack of ordering in 
quantum rotor models under aforementioned conditions.
 
Besides the fact that the lack of Bose-Einstein condensation has been proven in continuous systems in $d=1$ and $d=2$ more than 40 years ago, 
this is not the case of {\em lattice} models of interacting bosons. Canonical example of such lattice models is the {\em Bose-Hubbard} one. 
The fermionic Hubbard model is widely known from early sixties of last century \cite{Hubbard}. Its bosonic version was much less famous up to eighties, 
where it became popular after the paper \cite{Fishery}. In this paper basic features of these model have been examined. A breakthrough in an interest 
to the model dates about ten years ago, when it was applied to description of  trapped atomic gases and Bose-Einstein condensation therein  
\cite{GazyAtomowe}. Although there is a broad  `folk knowledge' that there is no Bose-Einstein condensation at positive temperatures in 
dimensions one and two, we couldn't find -- perhaps surprisingly -- a proof of this fact. This opportunity motivated us to  fill this gap.
 
Two problems appear here. One is the formal calculation, based on the Bogolyubov inequality. We define the order parameter to be 
the average of zero-momentum annihilation operator (like in \cite{BouzianeMartin} and other papers). By a suitable choice of 
operators one shows that this order parameter is zero in dimensions one and two 
at positive temperatures. 

The second aspect 
is justification of these calculations;
the ordinary (finite dimensional) Bogolyubov inequality is not sufficient here. 
It turned out that only in limited range one can use the technique developed 
in the paper \cite{BouzianeMartin}.
Instead, we consider certain sequence of finite-dimensional
approximations together with taking the infinite-dimensional limits 
on the level of thermal averages  and showing that these limits exist. As a byproduct, we obtain some result concerning density: We prove that it is bounded and nonzero for all values of parameters. For this result,
obvious from physical point of view, we couldn't find -- again surprisingly --  the proof in literature.
 
The organization of the paper is as follows. In  Sec. \ref{sec:setup} we present a `setup', i.e. the notation, formalism and  definition   
of the Bose-Hubbard model. This Section contains also some preparatory theorems, necessary for further considerations. 
In  Sec. \ref{sec:Bogolub}
 we elaborate some general aspects of Bogolyubov inequality necessary in further considerations.
 Sec. \ref{sec:BHandBIneq} presents the choice of operators in the Bogolyubov inequality and calculation of necessary commutators. 
 Sec.~\ref{sec:TDLim} describes  passing to the thermodynamic limit  to prove the lack of the Bose-Einstein condensation in 
dimensions one and two at positive temperatures. We give here also the proof of estimations for density.  Sec. \ref{sec:Summary} is devoted to summary, conclusions and some open (as far as we know) problems.

\section{Setup for the Bose-Hubbard model: definitions and preparatory theorems}
\label{sec:setup}
In this section we introduce the model and prove self-adjointness of its Hamiltonian.
Let $\cal{K}$ be a separable Hilbert space and $(e_0,e_1,\dots)$ an orthonormal basis; $e_n$ will also be denoted by $|n\kket$. 
Let $c^\dg, c$ be the standard  creation and anihilation operators on $\cal{K}$ with respect to the orthonormal basis $(e_n= |n\kket)$:
\be
c^\dg |n\kket := \sqrt{n+1} |n+1\kket,\,n\geq 0\,; \quad c|0\kket := 0 , \quad c|n\kket := \sqrt{n} |n-1\kket\,,\,n>0
\ee
and extended by linearity to the space of {\em finite linear combinations} of basis vectors.
Clearly on this space they satisfy
$\displaystyle  [c, c^\dg] =1 $ and  for the number operator $\hat{n}:= c^\dg c$ we have  
$\displaystyle  \hat{n} |n \kket = n|n\kket.$
For further applications, let us note, that since $\dsp (c\pm I)^\dg (c\pm I)\geq 0$,  we have the estimate 
\begin{equation}\label{c-oszac}
-(\hat{n}+I)\leq \,c^\dg + c\, \leq (\hat{n}+I)
\end{equation}
For a {\em finite} set $\La$ we define 
\begin{equation}
\Hcal_\La:= \bigotimes_{\xgr\in\La} H_\xgr, \quad {\rm where}\quad  H_\xgr:=\cal{K}.\end{equation}
Vectors  of the  induced orthonormal basis in $\Hcal_\La$  will be abbreviated as:
\begin{equation}\label{baza}
|n_1, n_2,\dots, n_{|\La|}\kket := |n_1\kket\otimes | n_2\kket\otimes \dots \otimes |n_{|\La|}\kket
\end{equation}
Let $\DD$ be the space of {\em finite} linear combinations of elements of basis in $\Hcal_\La$:
\begin{equation}\label{def-D}
\DD:=span \left \{ |n_1,\dots, n_{|\La|}\kket\,:\,n_1,\dots, n_{|\La|}\in \Ndb\cup\{0\}\,\right\},
\end{equation}
clearly $\DD$  is dense in $\Hcal_\La$.

Let us  define several linear operators acting on $\DD$.
For $\xgr\in\La$  by $c^\dg_\xgr$, $c_\xgr,\hat{n}_\xgr$ we denote linear operators acting as $c^\dg$, $c, \hat{n}$  
on  $\xgr$-th ''slot'' and as $1$ on remaining ``slots'' i.e.
\be 
\begin{split}
c^\dg_\xgr|n_1,\dots, n_{\xgr},\dots, n_{|\La|}\kket &:= \sqrt{n_{\xgr}+1}\, |n_1,\dots, n_{\xgr}+1,\dots, n_{|\La|}\kket\\
c_\xgr|n_1,\dots, n_{\xgr},\dots, n_{|\La|}\kket &:= \sqrt{n_{\xgr}} \, |n_1,\dots, n_{\xgr}-1,\dots, n_{|\La|}\kket\\
\hat{n}_\xgr|n_1,\dots, n_{\xgr},\dots, n_{|\La|}\kket &:= n_{\xgr} \, |n_1,\dots, n_{\xgr},\dots, n_{|\La|}\kket,
\end{split}
\ee
clearly $\displaystyle \hat{n}_\xgr=c^\dg_\xgr c_\xgr$. It is straightforward to check  that these operators, as operators on ${\cal{D}}$,  satisfy
\begin{align}\label{basic-rel}
[c_\xgr\,,\,c^\dg_\ygr]&=\delta_{\xgr\ygr} \,,& \, [\hat{n}_\xgr\,,\,c_\ygr] &= -\delta_{\xgr\ygr}  c_\ygr\,,\nonumber\\
 [\hat{n}_\xgr\,,\,c^\dg_\ygr] &= \delta_{\xgr\ygr}  c^\dg_\ygr \,,&\, [\hat{n}_\xgr\,,\,c^\dg_\ygr c_\zgr] &= (\delta_{\xgr\ygr} -\delta_{\xgr\zgr}) c^\dg_\ygr c_\zgr
\end{align}
\begin{remark}
It is known that $c_\xgr$ and $c^\dg_\xgr$ are closable and $c^\dg_\xgr=(c_\xgr)^*|_\DD$. 
In what follows for an linear operator $A$ acting on $\DD$ and such that $\DD\subset D(A^*)$ we will often use notation 
$A^\dg:=(A^*)|_\DD$.
\end{remark}

\noindent Notice that for a complex number $t_{\xgr\ygr}$ with polar decomposition $t_{\xgr\ygr}=e^{2 i \theta}|t_{\xgr\ygr}|$ from the obvious inequality 
$$\left (e^{- i \theta} \sqrt{|t_{\xgr\ygr}|}\, c_\xgr \pm e^{i \theta} \sqrt{|t_{\xgr\ygr}|}\, c_\ygr \right)^\dg 
\left (e^{- i \theta} \sqrt{|t_{\xgr\ygr}|} \, c_\xgr \pm e^{i \theta} \sqrt{|t_{\xgr\ygr}|} \,c_\ygr \right)\geq 0$$ 
we  get estimates:
\begin{equation}\label{c-T-oszac}
-|t_{\xgr\ygr}| (\hat{n}_\xgr+ \hat{n}_\ygr)\leq \, t_{\xgr\ygr} c_\xgr^\dg c_\ygr +  \overline{t_{\xgr\ygr}}  c_\ygr^\dg c_\xgr\,\leq \,
|t_{\xgr\ygr}| (\hat{n}_\xgr+ \hat{n}_\ygr).
\end{equation}
\noindent
The total number operator $\widehat{N}_\La$ and the operator $\widehat{N}_{2,\La}$ are defined as 
\be \widehat{N}_\La:=\sum_{\xgr\in \La} \hat{n}_\xgr\,,\,\quad\quad \widehat{N}_{2,\La}:=\sum_{\xgr\in \La} \hat{n}_\xgr^2
\ee
so that 
\be 
\begin{split}
\widehat{N}_\La |n_1,\dots, n_{|\La|}\kket &= (n_1 + \dots + n_{|\La|})  |n_1,\dots, n_{|\La|}\kket\\
\widehat{N}_{2,\La}  |n_1, \dots, n_{|\La|}\kket &= (n_1^2 + \dots + n_{|\La|}^2)  |n_1, \dots, n_{|\La|}\kket
\end{split}
\ee
If it doesn't lead to any confusion we will omit subscript $\La$ and denote these operators by $\widehat{N}$ and $\widehat{N}_2$. 
For $m, M \in\Ndb\cup\{0\}$ let 
\begin{equation}\label{def-DM}
\begin{split}
\DD_m &:= span\{ | n_1,\dots, n_{|\La|}\kket\,: n_1+ \dots + n_{|\La|}=m\}\\
\DM{M} &:= span\{ | n_1,\dots, n_{|\La|}\kket\,: n_1+ \dots + n_{|\La|}\leq M\}=\bigoplus_{m=0}^M {\cal{D}}_m \end{split}
\end{equation}
i.e. $\DD_m$ is the eigenspace of $\widehat{N}$ with eigenvalue $m$. Using this notation we have $\displaystyle \Hcal_\La=\bigoplus_{m=0}^\infty \Dm$ 
(orthogonal direct sum).  Moreover
\begin{equation}\label{monoton}
M < L  \Rightarrow \DM{M}\subset \DM{L} \,\,\,\quad{\rm and }\,\,\,\,{\cal{D}}=\bigcup_{M\in\Ndb} \DM{M}
\end{equation}
It is routine to check  that operators $\widehat{N}$ and  $\widehat{N}_{2}$ are essentially self-adjoint; their self-adjoint closures will be denoted by 
$\widehat{\NN}$ and $ \widehat{\NN}_2$, respectively. {\em Note the type of font, we will use similar notation for other operators as well. 
In general, the closure of an operator $A$ will be denoted  by $\overline{A}$ as well as complex conjugation of a number $z$ by $\overline{z}$.}

\noindent
We define two more operators on $\DD$:
\be\label{def-T}
T_\La:=\sum_{\xgr,\ygr\in\La} t_{\xgr\ygr} c^\dg_\xgr c_\ygr=\sum_{\xgr\neq \ygr} t_{\xgr\ygr} c^\dg_\xgr c_\ygr + \sum_{\xgr} t_{\xgr\xgr} \hat{n}_\xgr=: T'+ T'' 
\,,\,\, t_{\xgr\ygr}=\overline{t_{\ygr \xgr }}\in\Cdb
\ee
\be \label{def-L} L_\La :=\sum_\xgr(c^\dg_\xgr + c_\xgr). \ee
Let us note that operators $T', T'', \widehat{N}_2$ and $L$ are {\em symmetric} on ${\cal{D}}$ and 
\begin{equation}\label{zawieranie}
T'\Dm\subset \Dm\,,\,T''\Dm\subset \Dm\,,\, \widehat{N}_2\Dm\subset \Dm\,,\, L {\cal D}_m \subset {\cal D }_{m-1} \oplus {\cal D}_{m+1} 
\end{equation}

Finally, for  $\UU>0, \mu, \lambda\in \Rdb$,  let us  consider the {\em grand canonical ensemble}  hamiltonian:
\be
H_\La(\UU) := \UU \,\hat{N}_{2,\La}  +T_\La  - \mu\, \hat{N}_\La + \lambda L_\La  
\label{HamWZK}
\ee
(again we will omit subscript $\La$ in  $T_\La$ and $L_\La$). Of course $H_\La$ depends of $\lambda$ and $\mu$ as well, 
but we will use  dependence of $\UU$ explicitly, so we underline it in notation. 

\noindent
The operator $T_\La$ possess physical interpretation as the lattice {\em hopping term}, and $\UU \hat{N}_{2,\La} $ is the {\em on-site interaction term}.
We have also the  $L$ term, which is responsible for breaking of the  $U(1)$ symmetry. Realize that in the canonical ensemble
the average of the operator 
$\sum_\xgr c^\dg_\xgr$ (as well as $\sum_\xgr c_\xgr$) is equal to zero. However, 
in the grand canonical ensemble both averages can be non-zero. 
One takes (suitably scaled) averages of the operator 
$\sum_\xgr c^\dg_\xgr$ (or $\sum_\xgr c_\xgr$) as the `order parameter', i.e. condensate density.

Let us also fix two  numbers $\MM, \MM_d$ satisfying:
\begin{equation}\label{def-M}
\max \left\{\,\sum_{\ygr\in\La} | t_{\xgr\ygr} | \,,\,\, \xgr \in \Lambda\,\right\}\leq \MM \,,\quad \max\{|t_{\xgr\xgr}|:\xgr\in\La\}\leq  \MM_d
\end{equation}
\begin{remark}\label{stale-M}
This is not a restriction on {\em a single system}, but later on, in  termodynamic limit, 
we shall assume that $\MM$ and $\MM_d$ fulfilling (\ref{def-M}) can be chosen independently of $\Lambda$. Clearly we may assume $\MM_d\leq \MM$.
\end{remark}

\noindent
{\em At this moment we don't impose any other restrictions on the model; they will appear later on.} 
The rest of this section is devoted to the proof of self-adjointness of the Hamiltonian (\ref{HamWZK}).
\begin{theorem}\label{tw1}  
Let $\UU>0$ and  $\mu,\,\lambda\in \Rdb$.  Then
\begin{itemize}
\item[i)] The operator $H(\UU)$ defined by (\ref{HamWZK}) is essentially self-adjoint and its closure $\pH(\UU)$ is equal to
$\pH(\UU)=\UU \widehat{\NN}_2 +\overline{T  - \mu\, \hat{N} + \lambda L}$ (in particular this equality means that $D(\pH(\UU))=D( \widehat{\NN}_2)$).\vsm
\item[ii)] $\pH(\UU)$ is bounded from below  with lower bound $\gamma(\UU)$ satisfying \vsm
\begin{equation}\label{H-bound}
\gamma(\UU)\geq - K\, \frac{|\Lambda|(\MM/2+ 2 |\lambda|) +(\MM_d+|\mu|)}
{1- \frac{|\Lambda|}{\UU K }\left(|\Lambda|(\frac{\sqrt{3}}{2}\MM+ 2 |\lambda|) + \MM_d+|\mu|\right)}
\end{equation}
for any $\displaystyle K\in\Ndb\,\,{\rm such\,that}\,\,K > \frac{|\Lambda|}{\UU }\left(|\Lambda|(\frac{\sqrt{3}}{2}\MM+ 2 |\lambda|) + \MM_d+|\mu|\right)$
\item[iii)] The operator $\exp(-\beta \pH(\UU))$ is trace class for every $\beta>0$\vsm
\end{itemize}
\end{theorem}
The idea of the proof is to show that the hamiltonian (\ref{HamWZK}) 
is a ``small'' perturbation of the operator $\UU\widehat{N}_2$  and then to use  some general results  of perturbation theory -- they are   
summarized in Prop.~\ref{TW-pomoc}. After the proof of this proposition, which essentially consists of pointing to relevant results from literature, 
we prove -- in Prop.~\ref{prop-N2-estimates} -- estimates that enable us to apply Prop.~\ref{TW-pomoc} to the Bose-Hubbard Model.

\subsection{Some  results about unbounded perturbations.} 
Here we deal with unbounded perturbation and for completness of exposition we recall the following 
\begin{defi} \label{T-bounded}(\cite{Kato}, Ch.~5, 4.1)
{\em Let $T$ and  $A$ be operators acting on a Hilbert space}. The operator  $A$ is $T$-bounded {\em iff  $D(T)\subset D(A)$ and 
there exist $a,b\in\Rdb_+\cup\{0\}$  such that for every $\displaystyle \psi\in D(T):$
\begin{equation} \,||A\psi|| \leq a ||\psi||+ b||T\psi||.   \label{T-bounded-ineq1}\end{equation}
In this situation we will also say that}  $A$ is $T$-bounded with constants $(a,b)$. 

\noindent
{\em $A$ is $T$-bounded} with the relative bound $0$ {\em
if for every  $\epsilon>0$ there exist $a$ such that $A$ is $T$-bounded with constants $(a,\epsilon)$.}
\end{defi}
\begin{remark} There is an equivalent definition (\cite{Kato}, Ch.~5, 4.2) in which inequality (\ref{T-bounded-ineq1}) is replaced by 
\begin{equation} \,||A\psi||^2 \leq \tilde{a}^2 ||\psi||^2 + \tilde{b}^2||T\psi||^2.
\label{T-bounded-ineq2}\end{equation}
Indeed,  if (\ref{T-bounded-ineq2}) is satisfied then also (\ref{T-bounded-ineq1}) holds  with $a=\tilde{a}$ and $b=\tilde{b}$. In the opposite direction, 
if (\ref{T-bounded-ineq1}) holds then taking $\tilde{a}:=a\sqrt{1+1/\epsilon}$ and $\tilde{b}:=b\sqrt{1+\epsilon}$ for any $\epsilon>0$ we obtain  
(\ref{T-bounded-ineq2}). 
\end{remark}
\begin{remark} \label{rem-trans} From the very definition it follows that relative boundedness is transitive: 
if $A$ is $T$-bounded with constants $(a_1, b_1)$ and $T$ is $S$-bounded with
constants $(a_2,  b_2)$ then $A$ is $S$-bounded with constants $(a_1+b_1 a_2, b_1 b_2)$.
\end{remark}
Now we can formulate the proposition being the main tool of our analysis. 
\begin{proposition} \label{TW-pomoc} Let $\pT$ be a self-adjoint operator acting on a Hilbert space ${\cal H}$. Let  ${\cal D}\subset {\cal H}$ be a dense linear space and
$A: \DD\rightarrow {\cal H}$ be a symmetric operator.
Assume that
\begin{enumerate}
\item[i)] $\pT$ is essentially self-adjoint on $\DD$;\vsm
\item[ii)] The operator $A$ is $\pT|_\DD$-bounded with constants $(a,b)$ and  $b<1$.
\item[iii)] $\pT$ is bounded from below with the lower bound $\gamma_T$;\vsm
\item[iv)] $e^{-\beta \pT}$ is a trace class operator for every $\beta>0$\vsm
\end{enumerate}\vsm\vsm
Then 
\begin{enumerate}
\item The operator $\pT|_\DD+A$ is essentially self-adjoint and $\pS:=\overline{(\pT|_\DD+A)}=\pT+\overline{A}$ is self-adjoint on $D(\pT)$.
\item The self-adjoint operator $\pS$ is  bounded from below with the lower bound $\gamma_S$ and
\begin{equation}\label{gamma-bound} \gamma_S \geq \gamma_T-\max \left\{\frac{a}{1-b}\,, \,a+b |\gamma_T|\right\}=:\tilde{\gamma}_T\end{equation}
\item   The operator $e^{-t\pS}$ is a trace class operator for any $t>0$.
\end{enumerate}
\end{proposition}
\begin{remark} \label{remark-TW-pomoc}  If a pair $(\pT,A)$ fulfills assumptions $i)-iv)$ with constants $(a,b,\gamma_T)$ then for $q>0$ 
and $0\leq p\leq q$ the 
pair  $(q \pT, p A)$ fulfills  $i)-iv)$ with constants $(p a, \frac{p}{q} b , q \gamma_T)$.\\
\end{remark}\vspace{-2ex}

\noindent
{\em Proof of Prop.~\ref{TW-pomoc}:}

\noindent
{\bf 1)}  This statement is just the  Thm 4.4, Ch.5 of \cite{Kato} (Kato-Rellich Theorem) 
applied to operators $\pT|_{\cal D}$ and $A$;  only  assumptions $i)$ and $ii)$ are used.


\noindent
{\bf 2)} From the proof of mentioned theorem, it follows that $\overline{A}$ is $\pT$-bounded with (the same) constants $(a,b)$. 
The statement (2) is the  Thm 4.11, Ch.5  of \cite{Kato} applied to the pair $(\pT, \overline{A})$;

\noindent
{\bf 3)} Let us notice that, by the remark~\ref{remark-TW-pomoc}, 
it is enough to prove the third claim  for $t=1$. Therefore we shall prove that $e^{-\pS}$ is a 
trace class operator. To this end we are going to use the following 

\begin{lemma} {\rm ( \cite{Ruskai}, Thm.~4)}\label{Ruskai4}
Let $A, B$ be self-adjoint operators and  ${\cal D} \subset D(A)\cap D(B)$ 
be a  dense linear space. Assume that  $\Tr(e^{-A}) <\infty$, $B$ is bounded from 
below and $S:=\overline{(A+B)|_{\cal D}}$ is self-adjoint.  Then  $\Tr(e^{-S})\leq \Tr(e^{-A}e^{-B})$ and therefore  $e^{-S}$ is trace 
class. \hspace{\fill}$\Box$\hspace*{1em}
\end{lemma} 

Let us take $\alpha$ satisfying $b<\alpha<1$ and consider operators $\alpha \pT$ and $A$. 
These operators satisfy assumptions of the theorem with the same ${\cal D}$ and 
constants $\gamma_1:=\alpha \gamma_T$, $a_1:=a$ and $b_1:=\frac{b}{\alpha}< 1$. Therefore we know, by 1) and 2), 
that the operator $\pS_1:=\alpha \pT + \overline{A}$ 
is self-adjoint (on $D(\pT)$) and ${\cal D}$ is a core for $\pS_1$. It is also bounded from below with the bound 
$$\gamma_{S_1}\geq \alpha\gamma_T-\max \left\{\frac{a\alpha }{\alpha-b}\,, \,a+b |\gamma_T|\right\}$$
Therefore we can write the operator $\pS$ as a sum of two self-adjoint, bounded from below operators:
\begin{equation}\label{alpha-rozklad}
\pS=\pT+\overline{A}=(1-\alpha) \pT + ( \alpha \pT+ \overline{A}),
\end{equation}
where  $\exp(-(1-\alpha)\pT)$ is trace class (by $iii)$) and ${\cal D}$ is a core for $\pS$. 
Now by Lemma~\ref{Ruskai4} the operator $\exp(-\pS)$ is  trace class.  \hspace*{\fill} $\Box$
\vs

\begin{cor}\label{wniosek1}
Let $\pT$ and $A$ satisfy assumptions $i)-iii)$ of  Prop.~\ref{TW-pomoc} and let $B$ be  $\pT$-bounded with constants $(\tilde{a},\tilde{b})$. 
Then, for any $\beta>0$,  the operator $B\exp(-\beta(\pT+\overline{A}))$ is bounded and
\begin{equation}\label{wniosek1-1}
\begin{split}
||B\exp(-\beta(\pT+\overline{A})) ||\leq  
 (\tilde{a}+ \frac{a \tilde{b}}{1-b}) \,\exp(-\beta\tilde{\gamma}_T) 
+ \frac{ \tilde{b}}{1-b}\,\max\left\{|\tilde{\gamma}_T| \,\exp(-\beta\tilde{\gamma}_T), \frac{1}{e\beta}\right\},
\end{split}
\end{equation}
where $\tilde{\gamma}_T$ is defined in (\ref{gamma-bound}). 
\end{cor}

Let us remark  that the right hand side of the inequality depends on $A$ only through constants $a$ and $b$.\\
{\em Proof:}  Let us denote $\pS:=\pT+\overline{A}$. By the functional calculus of self-adjoint operators (see e.g.  \cite{Dun-Sch}, XII.2.7(c))  
we know that $\exp(-\beta \pS) : \Hcal\rightarrow D(\pS)=D(\pT)\subset D(B)$, therefore 
$B\exp(-\beta \pS )$ is defined on the whole space $\Hcal$ and  for $\varphi\in\Hcal$:
\begin{equation}\label{wniosek1-2}
||B \exp(-\beta\pS ) \varphi||
\leq \tilde{a} ||\exp(-\beta \pS)  \varphi|| + \tilde{b} ||T \exp(-\beta \pS) \varphi||
\end{equation}
For $\psi\in D(\pT)=D(\pS)\subset D(A)$ we have:
$$||\pT\psi|| \leq ||\pS \psi||+ ||\overline{A} \psi|| \leq  ||\pS \psi|| +   a ||\psi||+ b ||\pT\psi||,$$
and,  since $b<1$, 
$$||\pT\psi||\leq \frac{a}{1-b} ||\psi| + \frac{1}{1-b}||\pS\psi||$$
Using this estimate in (\ref{wniosek1-2}) for $\psi:=\exp(-\beta \pS)  \varphi$ we obtain: 
$$||B\exp(-\beta\pS) ||\leq  (\tilde{a}+ \frac{a \tilde{b}}{1-b}) \,||\exp(-\beta \pS)||  
+ \frac{ \tilde{b}}{1-b}\,||\pS \exp(-\beta\pS)||
$$
By the functional calculus and estimate (\ref{gamma-bound}) we have  inequalities $||\exp(-\beta \pS)||\leq \exp(-\beta\tilde{\gamma}_T)$ and 
$||\pS \exp(-\beta\pS)||\leq \max\left\{|\tilde{\gamma}_T|\, \exp(-\beta\tilde{\gamma}_T), \frac{1}{e \beta}\right\}$.
The formula  (\ref{wniosek1-1}) follows.
\hspace*{\fill}$\Box$\hspace*{1em}\vs

\subsection{Self-adjointness of Bose-Hubbard Hamiltonian}
To prove Thm.~\ref{tw1},  we  use Prop.~\ref{TW-pomoc} with $\pT:=\UU\widehat{\NN}_2$ and $A:=T  - \mu\, \hat{N} + \lambda L$. 
It is clear that $\widehat{\NN}_2\geq 0$ and it is easy to check  that $\exp(-\beta \widehat{\NN}_2)$ is  trace class for every $\beta >0$, 
so the same is true  for $\exp(-\beta \UU \widehat{\NN}_2)$.
Thus to prove the theorem we need to show that the operator $T  - \mu\, \hat{N} + \lambda L$ is 
$\UU\widehat{N}_2$-bounded with some constants $(a,b)\,,\,b<1$. In fact we are proving in Prop.~\ref{prop-N2-estimates} that each of operators 
$T'$,  $T''$, $\widehat{N}$ and $L$ is  $\widehat{N}_2$-bounded {\em with relative bound $0$}.  
Notice that this is enough,  since  it is  straightforward to see that then, for any $\lambda$ and $\mu$, 
the operator $T  - \mu\, \hat{N} + \lambda L$ is $\widehat{N}_2$-bounded with relative bound $0$; moreover from the very definition, 
it follows that if some  operator $A$ is $\widehat{N}_2$-bounded with relative bound $0$, it is also $\UU \widehat{N}_2$-bounded with relative bound $0$ for any 
$\UU\in\Rdb$.

\noindent
By  considerations above, the proof of the following proposition will complete the proof of 1) and 3) of  Thm.~\ref{tw1}.
\begin{proposition}\label{prop-N2-estimates}
For any $K\in \Ndb$ and any  $\psi\in {\cal{D}}$ we have the following estimates: 
\begin{equation}\label{T-prim-est}
||T'\psi||^2\leq \frac14  \MM^2 |\La|^2  (K+1)^2 ||\psi||^2 + \frac34  \frac{ \MM^2  |\La|^4 }{(K+1)^2} ||\widehat{N}_2 \psi||^2
\end{equation}
\begin{equation}\label{N-est}
||\widehat{N} \psi||^2 \leq K^2 ||\psi||^2 + \frac{|\Lambda|^2}{(K+1)^2} || \widehat{N}_2 \psi||^2
\end{equation}
\begin{equation}\label{T-bis-est}
||T''\psi||^2\leq \MM_d^2 K^2 ||\psi||^2 + \frac{ \MM_d^2 |\Lambda|^2}{(K+1)^2} || \widehat{N}_2 \psi||^2 
\end{equation}
\begin{equation}\label{L-est}
||L\psi|| \leq 2 |\Lambda| (K+1) ||\psi|| +\frac{2 |\Lambda|^2}{K+1}|| \widehat{N}_2 \psi||
\end{equation}
\end{proposition}
{\em Proof:}  Notice that, as mentioned above, these inequalities imply that operators $T',T'', \widehat{N}$ and $L$ are $\widehat{N}_2$-bounded with relative bound $0$.
Let us  start with the following
\begin{lemma} For any $\varphi\in \Dm$ we have:
\begin{equation}\label{T-prim-norma-Dm}
||T'\varphi||\leq  \MM \frac{m+1}{2} |\La| \,||\varphi||
\end{equation}
\begin{equation}\label{N-dwa-norma-Dm}
||\widehat{N}_2 \varphi||\geq \frac{m^2}{|\La|} ||\varphi|| 
\end{equation}
\end{lemma}
{\em Proof:} The first estimate we are going to prove is:
\begin{equation}\label{cxy-m-norm}
||c_{\xgr}^\dg c_{\ygr}\varphi||\leq \frac{m+1}{2}||\varphi||\quad {\rm for}\quad \xgr\neq \ygr \quad{\rm and }\quad \varphi\in \Dm
\end{equation}
Observe that each subspace $\Dm$ is an  orthogonal sum $\displaystyle \Dm=\bigoplus_{k=0}^{m} h_k$, 
where $ h_k:=span\{| n_1,\dots, n_{|\La|}\kket\in \Dm: n_\xgr+n_\ygr=k\}$.
We have $c_\xgr^\dg c_\ygr h_k\subset h_k$   and each $h_k$ is isomorphic to $h_k'\otimes h_k''$,  where
$$h_k':=span\{ | l, k-l\kket\,,\,l=0,\dots k\}\,,\quad h_k'':=span\{| n_1,\dots, n_{|\La|-2}\kket \,: n_1 +\dots + n_{|\La|-2}=m-k\}$$
and $c_\xgr^\dg c_\ygr (x_k'\otimes  x_k'')= A_k(x_k')\otimes x_k''$ where $$A_k|l,k-l\kket=\sqrt{k-l}\sqrt{l+1} |l+1,k-l-1\kket\,,\,l=0,\dots,  k $$
So in he basis $f_l:=|l,k-l\kket\,,\,l=0,\dots,  k$ the matrix of $A_k$ is 
$$A:=\left( \begin{array}{ccccc} 
0     &  0    &  \dots  &   \dots & 0\\
a_0   &  0    &   \dots &   \dots & 0\\
0     &  a_1  &   0     &   \dots & 0\\
\dots & \dots &   \dots &   \dots & \dots \\
0     &  0    &   \dots & a_{k-1} &  0 \end{array}\right)  \,, \quad  a_l:=\sqrt{l+1}\sqrt{k-l}$$
then $||A||^2=max\{ |a_l|^2: l=0, \dots,k-1\}$; this immediately follows from $||A||^2=||A^* A||$.
Therefore 
$$||A_k||^2= max \{(l+1)(k-l)\,,\, l=0,\dots, k-1\}=\left\{\begin{array}{l c} \left(\frac{k+1}{2}\right)^2 & k -  odd\\ \frac{k(k+2)}{4} & k - even\end{array}\right.,$$
and  
$$||(c_\xgr^\dg c_\ygr)|_{\Dm}||^2=max \{||A_k||^2\,,\,k=0,\dots,m\}\leq \left(\frac{m+1}{2}\right)^2$$
as in (\ref{cxy-m-norm}). Now, for  $\varphi\in \Dm$ we compute 
$$||T'\varphi||=||\sum_{\xgr\neq \ygr} t_{\xgr\ygr} c^\dg_\xgr c_\ygr \varphi||\leq \sum_{\xgr\neq \ygr} |t_{\xgr\ygr}| || c^\dg_\xgr c_\ygr \varphi||\leq \frac{m+1}{2}||\varphi||  \sum_{\xgr\neq \ygr} |t_{\xgr\ygr}|
\leq |\Lambda|\MM \frac{m+1}{2}||\varphi||$$
as claimed in (\ref{T-prim-norma-Dm}).\\
It remains to prove (\ref{N-dwa-norma-Dm}). This immediatly follows from the fact that the minimal value of $x_1^2+ x_2^2+\dots +x_l^2$ on the hyperplane $x_1+x_2+\dots+x_l=m$ is equal to 
$l \left(\frac{m}{l}\right)^2$. 
\hspace*{\fill}$\Box$\hspace*{1em}\vs

Now we can prove the proposition.
Let us fix $K\in \Ndb$. For any $\psi\in\cal{D}$, because of (\ref{monoton}) we may assume, that 
$\displaystyle \psi\in \DM{K+L}=\bigoplus_{m=0}^{K+L} \Dm$ for some integer $L>1$ and write $\displaystyle \psi=\sum_{m=0}^{K+L}\varphi_m\,,\,\varphi_m\in \Dm$ (orthogonal sum).
By (\ref{zawieranie}) and the estimates in the lemma we have:
\be
||T'\psi||^2=\sum_{m=0}^{K+L} ||T'\varphi_m||^2\leq \frac14 |\La|^2 \MM^2 \sum_{m=0}^{K+L} (m+1)^2 ||\varphi_m||^2
\ee
and
\be
\begin{split}
\sum_{m=0}^{K+L} (m+1)^2 & ||\varphi_m||^2 = ||\psi||^2 +\sum_{m=0}^{K+L} (m^2 +2 m) ||\varphi_m||^2= \\
&= ||\psi||^2 +\sum_{m=0}^K (m^2 +2 m) ||\varphi_m||^2 + \sum_{m=K+1}^{K+L} (m^2 +2 m) ||\varphi_{m}||^2\\
 & \leq ||\psi||^2+ (K^2 +2K) ||\psi||^2 + \sum_{m=K+1}^{K+L} (m^2 +2 m)\frac{|\La|^2}{m^4} ||\widehat{N}_2(\varphi_{m})||^2\leq\\
&\leq (K+1)^2 ||\psi||^2 +  |\La|^2  \left( \frac{1}{(K+1)^2} + \frac{2}{(K+1)^3}\right) \sum_{m=K+1}^{K+L}  ||\widehat{N}_2(\varphi_{m})||^2 \leq\\
&\leq (K+1)^2 ||\psi||^2 + |\La|^2  \left( \frac{1}{(K+1)^2} + \frac{2}{(K+1)^3}\right) ||\widehat{N}_2 \psi||^2 \leq \\
&= (K+1)^2 ||\psi||^2+   \frac{3 |\La|^2}{(K+1)^2} ||\widehat{N}_2 \psi||^2
\end{split}
\ee
Thus the  estimate (\ref{T-prim-est}) follows.
In the similar way:
\begin{equation}\begin{split}
||\widehat{N} \psi||^2 &= \sum_{m=0}^{K+L} m^2 ||\varphi_m||^2 = \sum_{m=0}^{K} m^2 ||\varphi_m||^2 +  \sum_{m=K+1}^{K+L} m^2 ||\varphi_m||^2 \leq \\
 &\leq  K^2 ||\psi||^2 + \frac{|\Lambda|^2}{(K+1)^2} \sum_{m=K+1}^{K+L} || \widehat{N}_2 (\varphi_{m})||^2
\leq K^2 ||\psi||^2 + \frac{|\Lambda|^2}{(K+1)^2} || \widehat{N}_2 \psi||^2
\end{split}
\end{equation}
This is the estimate (\ref{N-est}). Finally,  we compute:
\begin{equation}
\begin{split}
\left|(\psi \,|\, T''\psi)\right| &= \left|(\psi \,|\, \sum_{\xgr\in\La} t_{\xgr\xgr} \hat{n}_\xgr \psi)\right|\leq 
\sum_{\xgr\in\La} |t_{\xgr\xgr}| | (\psi \,|\,\hat{n}_\xgr \psi)|\leq \\
&\leq  \max\{|t_{\xgr\xgr}|:\xgr\in\La\} \sum_{\xgr\in\La} (\psi \,|\,\hat{n}_\xgr \psi)= \MM_d  (\psi \,|\, \hat{N}\psi),
\end{split}
\end{equation}
then, for $\varphi_m\in \Dm$ we get 
$$\left|(\varphi_m \,|\, T''\varphi_m)\right|\leq \MM_d (\varphi_m \,|\, \hat{N}\varphi_m)=  \MM_d  m ||\varphi_m||^2$$
therefore  
\begin{equation}\label{T-bis-norma-Dm}
||T''|_{\Dm}||\leq \MM_d m\,,  \quad {\rm and}
\end{equation}
\begin{equation*}\begin{split}
||T''\psi||^2=\sum_{m=0}^{K+L} ||T''\varphi_m||^2&\leq \MM_d^2 \sum_{m=0}^{K+L} m^2 ||\varphi_m||^2 = \MM_d^2||\widehat{N}\psi||^2\leq\\
&\leq \MM_d^2\left( K^2 ||\psi||^2 + \frac{|\Lambda|^2}{(K+1)^2} || \widehat{N}_2 \psi||^2\right), \end{split}
\end{equation*}
where in the last step we use second claim of the proposition, and this is (\ref{T-bis-est}).\\
Finally to prove the last inequality consider the lemma 
\begin{lemma}\label{c-cplus-est}
For every $\psi\in\DD$ the following inequalities hold 
\begin{equation*}
||c_\xgr\psi||\leq ||\widehat{\NN}^{1/2}\psi||,\quad ||c^\dg_\xgr\psi||\leq ||(\widehat{\NN}+1)^{1/2}\psi||
\end{equation*}
and, consequently, operators $\displaystyle c^\dg_{\xgr} (\widehat{\NN}+1)^{-1/2}$ and $\displaystyle c_{\xgr} (\widehat{\NN}+\rho)^{-1/2}$ extend to  bounded operators 
with the norm $\leq 1$ for any $\rho>0$.
\end{lemma}
{\em Proof:} This immediately follows from definitions of $c_\xgr, c^\dg_\xgr$ and $\widehat{\NN}$. \hspace*{\fill}$\Box$\hspace*{1em}

This way we obtain 
\begin{equation}\label{L-N-est}
||\sum_\xgr c_\xgr^\dg(\widehat{\NN}+1)^{-1/2}||\leq |\Lambda|\,,\quad||L(\widehat{\NN}+1)^{-1/2}||\leq 2 |\Lambda|,
\end{equation}
and 
$$ ||L\psi||=||L(\widehat{\NN}+1)^{-1/2}(\widehat{\NN}+1)^{1/2}\psi||\leq 2|\Lambda|||(\widehat{\NN}+1)^{1/2}\psi||\leq 2 | \Lambda|(||\widehat{N}\psi ||+ ||\psi||)$$
and (\ref{L-est}) follows now from (\ref{N-est}). The proposition is proven as well as points $1)$ and $3)$ of Thm.~\ref{tw1}. 

It remains to prove the inequality (\ref{H-bound}) in $2)$. By Prop.~\ref{prop-N2-estimates}:
$$||(T  - \mu\, \hat{N} + \lambda L)\psi||\leq a(\UU)||\psi||+ b(\UU)||\UU \widehat{N}_2 \psi||\quad,\quad\psi\in\DD,$$
where 
$$a(\UU):=(K+1) \left[|\Lambda|(\MM/2+ 2 |\lambda|) +(\MM_d+|\mu|)\right]$$
$$b(\UU):=\frac{|\Lambda|}{\UU(K+1)}\left(|\Lambda|(\frac{\sqrt{3}}{2}\MM+ 2 |\lambda|) + \MM_d+|\mu|\right).$$
Remembering that  $\UU\widehat{N}_2\geq \gamma= 0$, we have 
$\displaystyle\gamma(u)\geq -\frac{a(\UU)}{1-b(\UU)}$  by (\ref{gamma-bound}), and the inequality (\ref{H-bound}) follows.
\hspace*{\fill}$\Box$\hspace*{1em}

\noindent
By  the proof of Thm.~\ref{tw1}  we have:
\begin{cor}
For any $\omega \in \Rdb$ an operator  $\pH(\UU) +\omega I$ is self-adjoint, bounded from below and $\exp(-\beta(\pH(\UU)+\omega I))$ is trace class.
\end{cor}\hspace*{\fill}$\Box$\hspace*{1em}
\section{Bogolyubov inequality for systems of bosons}
\label{sec:Bogolub}
The fundamental technical tool in proving the absence of ordering is the 
{\em Bogolyubov inequality}. 
Working with bosons, one needs a kind of 
{\em infinite dimensional} version of that inequality involving unbounded operators. 
One approach is to obtain directly  such an inequality for a given hamiltonian 
and apprioprate class of operators -- 
this way for Bose gas in continuum was used in \cite{BouzianeMartin}.  
In our approach we proceed in different manner. Namely, we start with 
{\em finite dimensional} Bogolyubov inequality for finite dimensional ``approximations'' 
of relevant operators and show by 
limiting procedure that finite dimensinal averages converge to ``true'' averages for the model.

Let $V$ be a {\em finite dimensional} Hilbert space and $H=H^*$ a self-adjoint operator on $V$. For an operator $B$ on $V$ let
\begin{equation*}
\brak B \kket := \frac{\Tr_V(B e^{-\beta H})}{\Tr_V (e^{-\beta H})} \,,\,\beta>0.
\end{equation*}
Let $A,C$ be linear operators on $V$. The  Bogolyubov inequality \cite{Bogolubow62} reads: 
\be
\frac{\beta}{2} \brak A^* A + A A^*\kket \brak[[C,H], C^*]\kket \geq |\brak[C,A]\kket|^2.
\label{BogIneq}
\ee

Assume now, that $V$ is a subspace of a Hilbert space $\Hcal=V\oplus V^\perp$ and let $P_V$ be the orthogonal projection on $V$.
If $A$ is an operator on $V$ let $\tilde{A}:=A P_V$ be its extension (by $0$ on $V^\perp$) on $\Hcal$.
It is easy to see that for linear operators $A, A_1,\dots, A_k, H $ on $V$ we have 
$$\widetilde{A^*}=(\tilde{A})^*\,,\quad  \Tr_V ( A e^{-\beta H})=\Tr(\tilde{A} e^{-\beta \tilde{H}})\,,
\quad P(A_1,\dots, A_k)^{\sim}=P(\widetilde{A_1},\dots,  \widetilde{A_k}), $$
for any polynomial $P(x_1,\dots, x_k)$ of noncommuting variables. Thus the inequality (\ref{BogIneq}) can be written as 
\be
\frac{\beta}{2} \brak \tilde{A}^* \tilde{A} + \tilde{A} \tilde{A}^*\kket_V \brak[[\tilde{C},\tilde{H}], \tilde{C}^*]\kket_V 
\geq |\brak[\tilde{C},\tilde{A}]\kket_V|^2,
\ee
where, for operators $B:\Hcal \rightarrow \Hcal$ satisfying $B=BP_V=P_V B$,  we denote 
$$\brak B \kket_V := \frac{\Tr(B e^{-\beta \tilde{H}})}{\Tr(P_V e^{-\beta \tilde{H}})}$$

Finally, and this is the situation we deal with, let us assume  
that $\DD\subset \Hcal$ is a dense linear space with $V\subset {\cal D}$ and let $H, A, C: {\cal D}\rightarrow \Hcal$ be linear operators;
assume moreover that $H$ is symmetric and ${\cal D}\subset D(A^*)\cap D(C^*)$. Then restricting operators to $V$ and then extending to $\Hcal$ we obtain the following inequality:
\begin{equation}\label{finiteBogol}
\frac{\beta}{2} \brak (A_V)^* A_V + A_V (A_V)^*\kket_V \brak[[C_V,H_V], (C_V)^*]\kket_V 
\geq |\brak[C_V,A_V]\kket_V|^2,
\end{equation}
where $A_V:=P_V A P_V$, etc,  and the average $\brak B \kket_V$ for operators satisfying $B=B_V$ is defined this time  as
\begin{equation}\label{finiteAve}
\brak B \kket_V := \frac{\Tr(B e^{-\beta H_V})}{\Tr(P_V e^{-\beta H_V})}
\end{equation}

Using notation just introduced we can formulate:
\begin{proposition} \label{komut-finite} 
Let $\DD\subset \Hcal$ be  a dense linear space and  $H, A, C: {\cal D}\rightarrow \DD$ be linear operators;
assume moreover that $H$ is symmetric and ${\cal D}\subset D(A^*)\cap D(C^*)$. Let $V, W$ be finite dimensional subspaces satysfying $W\subset V\subset\DD$.\\
If $P_V C\subset CP_V$  then 
\begin{align} 
[C_V, A_V]&=[C,A]_V \label{komut-finite-1}\\
[[C_V, H_V], (C_V)^*]&=[[C,H],C^\dg]_V \label{komut-finite-2}
\end{align}
If, morever $P_V A\subset AP_W$ then additionally:
\begin{align} 
A_V(A_V)^*+ (A_V)^*A_V &=(A A^\dg +A^\dg A)_V - (P_V-P_W) A^\dg A (P_V-P_W) \label{komut-finite-3}
\end{align}
and 
\begin{equation}\label{finiteBogol1}
\frac{\beta}{2} \brak (A A^\dg +A^\dg A)_V\kket_V \brak[[C,H], C^\dg]_V\kket_V\geq |\brak[C,A]_V\kket_V|^2,
\end{equation}
\end{proposition}
\noindent{\em Proof: } It is straightforward to verify that for any operators $E, F : \DD\rightarrow \DD$:
\begin{equation}\label{kom-M}
[E_V, F_V]= [E,F]_V+ P_V E (P_VF - F P_V) P_V+ P_V F (EP_V - P_VE )P_V.
\end{equation}
If an  operator $P$ satisfies  $PP_V=P$ then $P (P_VE -E P_V) P_V =0.$
{\em In particular  if either $P_V E\subset EP_V$ or $P_V F\subset FP_V$  then  (\ref{kom-M}) reduces to $[E_V, F_V]= [E,F]_V$.} 
So  (\ref{komut-finite-1})  follows from (\ref{kom-M}) by assumptions on $C$.


To prove (\ref{komut-finite-2}) notice that  our assumptions on $C$ imply:
\begin{equation}\label{CPM-rel}
(C_V)^*=(C^\dg)_V=C^\dg P_V\supset P_V C^\dg,
\end{equation}
and  for any operator $E:\DD\rightarrow \DD$:
\begin{equation*}\begin{split}
[[C_V, E_V], (C^\dg)_V] =& P_V C P_V E P_V C^\dg P_V -  P_V E P_V C P_V C^\dg P_V +\\
&- C^\dg P_V C P_V E P_V + C^\dg P_V E P_V C P_V=\\
=& P_V C E C^\dg P_V   -  P_V E C C^\dg P_V- P_V C^\dg C E P_V  + P_V C^\dg  E C P_V=\\
=& P_V [[C, E], C^\dg]P_V,
\end{split}
\end{equation*}
In particular for $E=H$ we obtain (\ref{komut-finite-2}).


To prove  (\ref{komut-finite-3})  let us notice that 
\begin{equation}\label{APM-rel}
(A_V)^*=(A^\dg)_V= A^\dg P_V\supset P_W A^\dg \,\,,\, 
\end{equation}
Using these relations, let us compute 
\begin{equation*}
A_V(A_V)^*=P_V A P_V A^\dg P_V=P_V A P_V A^\dg P_V P_V=P_V A P_V P_W A^\dg P_V=P_V A P_W A^\dg P_V=P_V A A^\dg P_V
\end{equation*}
\begin{equation*}
(A_V)^* A_V =P_V A^\dg P_V A P_V=P_V A^\dg A P_W P_V= P_V A^\dg A P_W
\end{equation*}
and 
\begin{equation*}
\begin{split}A_V(A_V)^*+ (A_V)^*A_V &= P_V A A^\dg P_V+ P_V A^\dg A P_W= P_V (A A^\dg +  A^\dg A) P_V- P_V A^\dg A (P_V-P_W),\\
&= P_V (A A^\dg +  A^\dg A) P_V- (P_V-P_W) A^\dg A (P_V-P_W)
\end{split}\end{equation*}
The last equality follows from (\ref{APM-rel}): 
$$P_W A ^\dg A (P_V-P_W)\subset  A^\dg P_V A (P_V-P_W) \subset A ^\dg A P_W (P_V-P_W)=0$$

It remains to prove (\ref{finiteBogol1}),  in fact it follows from inequality (\ref{finiteBogol}). The RHS of (\ref{finiteBogol}) reads:
$$|\brak[C_V,A_V]\kket_V|^2=|\brak[C,A]_V\kket_V|^2 ,$$
and, using an abbreviation  $q_V:=P_V-P_W$,  the LHS (without $\frac{\beta}{2}$):
\begin{equation*}
\begin{split}
& \brak (A_V)^* A_V + A_V (A_V)^*\kket_V \brak[[C_V,H_V], (C_V)^*]\kket_V =\\ 
&=  \brak (A A^\dg +A^\dg A)_V\kket_V \brak[[C,H], C^\dg]_V\kket_V - \brak q_VA^\dg A q_V \kket_V \brak[[C_V,H_V], (C_V)^*]\kket_V 
\end{split}
\end{equation*}
It is known that $\brak[[C_V,H_V], (C_V)^*]\kket_V \geq 0$, (see e.g \cite{AsaAuer})  and  
$\brak q_VA^\dg A q_V \kket_V \geq 0 $ (as a thermal average of positive operator)
therefore inequality   (\ref{finiteBogol}) gives
\begin{equation*}
\frac{\beta}{2} \brak (A A^\dg +A^\dg A)_V\kket_V \brak[[C,H], C^\dg]_V\kket_V\geq |\brak[C,A]_V\kket_V|^2.
\end{equation*}
\hspace*{\fill}$\Box$\hspace*{1em}

We are going to use sequence of inequalities  (\ref{finiteBogol}) for  $V:=\DM{M}$ (defined in  (\ref{def-DM})). 
In this situation we will write  $P_M$ for  the orthogonal projection on $\DM{M}$, and 
for an operator $B: D(B)\rightarrow \Hcal$ with $\DD\subset D(B)\subset \Hcal$ we will write 
\begin{equation}\label{defBM}
B_M:= P_M B P_M\,,\quad\quad \brak B_M \kket_M := \frac{\Tr(B_M e^{-\beta H_M})}{\Tr(P_M e^{-\beta H_M})}
\end{equation}
For operators of interest  we will show  convergence of these finite dimensional approximations to their ``true'' thermal averages.

For a {\em bounded} operator $A$  its thermal average $\brak A \kket$ is defined as  
\begin{equation}\label{thermal}
\brak A \kket:=\frac{\Tr(A\exp(-\beta \pH))}{\Tr(\exp(-\beta \pH))},
\end{equation}
and for $\pH\equiv \pH(\UU)$ it is well defined and finite by Thm.~\ref{tw1}.
Let us note that thermal averages don't change under replacements $\pH\rightarrow \pH+\omega I$ for any $\omega\in\Rdb$. 
In the following, we have to  consider more general case of some {\em unbounded} observables $A$. For them we have to  show that the formula (\ref{thermal}) 
is meaningful.  More precisely we have:
\begin{proposition}\label{srednie}
Let  operators $\pT$ and $A$ satisfy assumptions $i)-iv)$ of Prop.~\ref{TW-pomoc} and  let $B$ be a $\pT$-bounded operator. 
Then for any $\beta>0$ the operator $B\exp(-\beta(\pT+\overline{A}))$ is trace class.
\end{proposition}

\noindent
{\em Proof:} Let $\pS:=\pT+\overline{A}$. By Cor.~\ref{wniosek1} the operator  $B\exp(-\beta \pS)$ is bounded. 
Since $B\exp(-\beta \pS)=\left(B\exp(-\frac{\beta}{2}\, \pS)\right) \exp(-\frac{\beta}{2}\, \pS)$ 
and, by Prop.~\ref{TW-pomoc} the operator $\exp(-\beta \pS)$ is trace class for every $\beta>0$,  the result is clear.
\hspace*{\fill}$\Box$\hspace*{1em}

Now we consider the problem of convergence. Our aim is to prove that for $B:\DD\rightarrow \Hcal$, possibly satisfying some additional conditions,
$\displaystyle \lim_{M\rightarrow\infty} \brak B_M \kket_M= \brak B \kket$.  We will prove this by replacing $\exp(-\beta \pH_M)$ in (\ref{defBM}) by a different 
operator, say $\exp(-\beta \tilde{\pH}_M)$, without changing the value of $\brak B_M \kket_M$ 
and the new sequence of operators will converge in $L^1(\Hcal)$ i.e. 
the Banach space of trace class operators on $\Hcal$.

\begin{proposition}\label{tw-conv-semigroups}
Let operators $\pT,A$, a linear space  $\DD$ and numbers $(a,b)$ satisfy assumptions of Prop.~\ref{TW-pomoc}. 
Let $(P_n)_{n\in\Ndb}$ be  an increasing  sequence of finite dimensional orthogonal projections  satisfying
\begin{equation}
\mbox{\rm strong\,-}\!\!\!\lim_{n\rightarrow\infty}P_n=I \quad,\quad P_n(\Hcal)\subset \DD\quad,\quad \DD\subset \bigcup_{n\in \Ndb} P_n(\Hcal)
\end{equation}
For   $\alpha$ satysfying  $1>\alpha > b$  and $\omega$ such that $\alpha \pT+ \overline{A}+\omega I\geq 0$ let us define, cf.  (\ref{alpha-rozklad}),   
a sequence of self-adjoint operators 
$\pS^\alpha_n$ on $D(\pS^\alpha_n):=D(\pT)$: 
\begin{equation}\label{def-Sn-alfa}
\pS^\alpha_n:=\left[(1-\alpha) \pT -\omega I\right]   + P_n ( \alpha \pT+ \overline{A}+\omega I)P_n =
\left[(1-\alpha) \pT -\omega I\right] + P_n ( \alpha \pT|_\DD+ A+\omega I)P_n.
\end{equation}

Then for any $t>0$ 
$$\lim_{n\rightarrow\infty} \exp\left[-t \pS^\alpha_n\right]= \exp\left[-t (\pT+\overline{A})\right]\quad {\rm in \,\, }L^1(\Hcal)$$
\end{proposition}

\noindent
{\em Proof:}  We will use the following 
\begin{lemma}\label{zagrebnov} {\rm [\cite{Zagr}, lemma p.~271] }
Let $(H_n)_{n\in \Ndb}$ be a sequence of self-adjoint operators and  $H_-, H$ be self-adjoint operators on a Hilbert space $\Hcal$. Assume that\\
1) $H_-\leq H_n\,\,\forall  n\in\Ndb $\\
2) $\exp(-\beta H_-)\in L^1(\Hcal)$ for any $\beta>0$\\ 
3) $\displaystyle \lim_{n\rightarrow\infty}H_n =  H$ in a strong-generalized sense.

\noindent
Then\\
1) $\exp (-\beta H)\in L^1(\Hcal)$ for any  $\beta> 0$\\
2) $\displaystyle \lim_{n\rightarrow\infty} \exp(-\beta H_n) = \exp (-\beta H)$  in $L^1(\Hcal)$. \hspace*{\fill}$\Box$\hspace*{1em}
\end{lemma}

Let us recall that the sequence of self-adjoint operators $T_n$ converges to a self-adjoint operator $T$ in 
{\em a strong generalized sense} iff for every $z\in\Cdb\setminus\Rdb$ the sequence
of resolvents $(T_n-zI)^{-1}$ converges strongly to $(T-zI)^{-1}$. 
The useful criterion for strong convergence of resolvents is:
\begin{lemma}\label{Kato-cor16} {\rm [\cite{Kato}, Ch.~VIII, Cor.~1.6]}
Let $(T_n)$ be a sequence of self-adjoint operators and $T$ a self-adjoint operator. 
Assume $\DD$ is a core for $T$ and $T_n \psi\rightarrow T\psi$ for every 
$\psi\in\DD$. Then for avery $z\in\Cdb\setminus\Rdb$ the sequence of resolvents $(T_n-zI)^{-1}$ converges strongly to $(T-zI)^{-1}$.
\hspace*{\fill}$\Box$\hspace*{1em}
\end{lemma}

Now, to prove our statement we use Lemma~\ref{zagrebnov}. We put   $H_-=(1-\alpha) \pT-\omega I $,  
$H_n :=\pS^\alpha_n$ and $H:=\pT+\overline{A}$. 
Clearly $\exp\left[-\beta ((1-\alpha) \pT-\omega I)\right]$ is trace class for any $\beta>0$ and  
$\pS^\alpha_n\geq (1-\alpha) \pT-\omega I$ due to the definition of $\omega$.   It 
remains to verify that $\pS^\alpha_n$ convereges to $\pT+\overline{A}$ in a  strong generalized sense. We use Lemma~\ref{Kato-cor16}.
For any $\psi\in\DD$,  
$P_n\psi=\psi$ for sufficiently large $n$, therefore
\begin{equation*}\begin{split} 
\lim_{n\rightarrow\infty} \pS^\alpha_n\psi &= \lim_{n\rightarrow\infty} \left[(1-\alpha) \pT-\omega I  + P_n ( \alpha \pT|_\DD+ A+\omega I)P_n\right]\psi =\\
&=(1-\alpha) \pT\psi -\omega \psi  + \alpha \lim_{n\rightarrow\infty} P_n \pT\psi+ \lim_{n\rightarrow\infty}  P_n A\psi +\omega \psi=
\pT\psi + A\psi =\\
&=(\pT + \overline{A}) \psi, 
\end{split}\end{equation*}
Since $\DD$ is a core for $\pT+\overline{A}$ the result follows. \hspace*{\fill}$\Box$\hspace*{1em}

\noindent
The next theorem is essential for convergence of finite dimensional approximations of thermal averages. In its proof we will use the following lemma:
\begin{lemma}\label{contra} Let $A$ be $T$-bounded with constants $(a,b)$. Then for  every contraction $P$ (i.e.  $||P||\leq 1$) satisfying $PT\subset TP$,  operators
$AP$ and $PAP$ are  $T$-bounded with (the same) constants $(a,b)$.
\end{lemma}

\noindent 
{\em Proof:} It is straightforward to check that if $P$ {\em is  any contraction} the operator $PA$ is $T$-bounded with (the same) constants $(a,b)$. 
So it is enough to prove that if $PT\subset TP$ then $AP$ is $T$-bounded with constants $(a,b)$. 
By assumptions we have $PD(T)\subset D(T)\subset D(A)$. Therefore for $\psi\in D(T)$: $T P\psi=P T\psi$ and 
$$||AP\psi||\leq a||P\psi||+ b||TP\psi||\leq a||\psi||+ b||PT\psi||\leq a||\psi||+b||T\psi||$$.
 \hspace*{\fill}$\Box$\hspace*{1em}

\begin{theorem}\label{tw-granica} 
Let us keep notation and assumptions of Prop.~\ref{tw-conv-semigroups}. In particular, $A$ is $\pT|_\DD$-bounded wich constants $(a,b)$.
We add two more assumptions: \\
1) $\quad b<\frac13\quad$ and  \quad \quad 2)  $ P_n \pT \subset \pT P_n \quad  {\rm for \,\, every}\,\,  n\in\Ndb$. \\
If  $B$ and $B^*$ are $\pT$-bounded, then
$$\lim_{n\rightarrow\infty} \Tr\left[P_nBP_n\exp(-\beta \pS^\alpha_n)\right]=\Tr[B\exp(-\beta (\pT+\overline{A}))]$$
\end{theorem}

\noindent {\em Proof:} Let  $\pS:=\pT+\overline{A}$. The operator 
$B\exp(-\beta \pS)$ is trace class by Prop.\ref{TW-pomoc} (3). 
We start by  proving that operators $B \exp(-\beta \pS^\alpha_n)\,,\,n\in\Ndb$ are  bounded for any $\beta>0$ (therefore also trace class) 
and give uniform (in $n$) bound for their norms. By (\ref{def-Sn-alfa}):
$$\pS^\alpha_n=\left[(1-\alpha) \pT-\omega I\right]  + P_n ( \alpha \pT|_\DD+ A+\omega I)P_n,$$
Let $\pT_1:=(1-\alpha) \pT -\omega I$ and $A_1:=\alpha \pT|_\DD+ A+\omega I$, then $\DD\subset D(A_1)$ and, for $\psi\in\DD$,:
\begin{equation*}\begin{split}
||A_1\psi||\leq & \alpha ||\pT\psi||+||A\psi||+ |\omega|\,||\psi||\leq (a+|\omega|)||\psi||+ (\alpha + b)||\pT\psi||=\\
&=(a+|\omega|)||\psi||+ \frac{\alpha + b}{1-\alpha}||(1-\alpha)\pT\psi||\leq 
\left(a+|\omega|\frac{1+b}{1-\alpha}\right)||\psi||+ \frac{\alpha + b}{1-\alpha}||\pT_1\psi||.
\end{split}\end{equation*}
By taking  $b<\alpha<\frac{1}{3}$ we get $\frac{\alpha + b}{1-\alpha}<1$. Therefore $\pT_1$ and $A_1$ satisfy assumptions $i)-iii)$  
of Prop.~\ref{TW-pomoc}  with constants $a_0:=a+|\omega|\frac{1+b}{1-\alpha}$ and $b_0:=\frac{\alpha + b}{1-\alpha}<1$. 
By the lemma~\ref{contra} $P_nA_1P_n$ are also 
$\pT_1$ bounded for any $n$ with the same constants.  On the other hand $B$ is $T_1$-bounded, since $T$ is $T_1$-bounded (c.f. remark \ref{rem-trans}).
By the corollary \ref{wniosek1} we get uniform bound on $||B \exp(-\frac{\beta}{2} \pS^\alpha_n )||$.

Now, we can write:
\begin{equation}
\begin{split}
P_n B P_n \exp(-\beta \pS^\alpha_n)  -  B \exp(-\beta \pS) = 
\Pi^1_n + \Pi^2_n,
\end{split}
\end{equation}
where
\[
\Pi^1_n \equiv \left[ P_n B P_n \exp(-\beta \pS^\alpha_n) -  B \exp(-\beta \pS^\alpha_n)\right] ,
\]
\[
\Pi^2_n
\equiv
\left[B  \exp(-\beta \pS^\alpha_n)- B  \exp(-\beta \pS)\right].
\]
The first term can be written as:
$$\Pi^1_n= -(1-P_n) B P_n  \exp(-\beta \pS^\alpha_n )- B(1-P_n) \exp(-\beta \pS^\alpha_n)$$

Note that $D(\pT_1(1-P_n))=D(\pT_1)$ and $\pT_1=(1-P_n)\pT_1(1-P_n)+ P_n \pT_1 P_n$. therefore 
\begin{equation}\label{PM-kom}
\begin{split}
\pS^\alpha_n &= P_n(\pT+\overline{A})P_n + (I-P_n) \pT_1 (I-P_n)  \,\,\,{\rm and}   \\
\pS^\alpha_nP_n & \supset P_n \pS^\alpha_n\,,\,\quad \exp(-\beta  \pS^\alpha_n)P_n = P_n \exp(-\beta  \pS^\alpha_n)
\end{split}
\end{equation}

To shorten notation let us denote $Q_n:=1-P_n$. By the second equality in (\ref{PM-kom}): 
$$ \Tr\left[Q_n B P_n  \exp(-\beta \pS^\alpha_n )\right]=\Tr\left[Q_n B P_n  \exp(-\beta \pS^\alpha_n )P_n\right]=0$$
and, again by (\ref{PM-kom}), 
$$Q_n \exp(-\beta \pS^\alpha_n)=Q_n \exp(-\beta \pS^\alpha_n) Q_n=Q_n \exp\left[-\beta \pT_1 \right] Q_n$$
So we have: 
\begin{equation}\begin{split}
\left|\Tr(\Pi^1_n)\right| &= \left|-\Tr\left[B Q_n \exp(-\beta \pS^\alpha_n)\right]\right|=\\
&= \left| \Tr\left[ B Q_n \exp\left(-\frac{\beta}{2} \pT_1 \right) Q_n\exp\left(-\frac{\beta}{2} \pT_1 \right) Q_n\right]\right|\\
&\leq 
\left|\left|Q_n B Q_n \exp\left(-\frac{\beta}{2} \pT_1 \right)\right|\right|\,\times \, 
\Tr \left[Q_n\exp\left(-\frac{\beta}{2} \pT_1 \right) Q_n\right]
\end{split}
\end{equation}
Since the operator $\exp\left(-\beta \pT_1 \right)$ is trace class: 
$$ \lim_{n\rightarrow \infty} \Tr\left[ Q_n  \exp\left[-\beta (1-\alpha) \pT \right] Q_n \right]=0,$$
and we obtain $\displaystyle \lim_{n\rightarrow \infty}\Tr(\Pi^1_n)=0$ {\em provided that} 
\begin{equation}\label{sup}
\sup_{n\in\Ndb} ||(1-P_n) B (1-P_n) \exp\left(-\frac{\beta}{2} \pT_1 \right)||< \infty.
\end{equation}
Since $(1-P_n)\pT\subset \pT (1-P_n)$ the operator $(1-P_n) B (1-P_n)$ is $\pT$-bounded by the lemma~\ref{contra}. 
Therefore, since $\pT$ is $\pT_1$ bounded,  also $\pT_1$-bounded (c.f. Remark \ref{rem-trans}),
with some constants $(a_0,b_0)$ independent of $n$.  
Thus, for $\psi\in D(\pT)$:
$$||(1-P_n) B (1-P_n) \exp\left(-\frac{\beta}{2} \pT_1 \right)\psi||\leq a_0||\exp\left(-\frac{\beta}{2} \pT_1 \right)\psi||+ 
 b_0 ||\pT_1 \exp\left(-\frac{\beta}{2} \pT_1 \right)\psi||$$
So the inequality (\ref{sup}) is true and  $\displaystyle\lim_{n\rightarrow\infty} \Tr(\Pi^1_n)=0$.

\noindent
Let us now consider $\Pi^2_n$:
\begin{equation*}
\begin{split}
B \exp(-\beta \pS^\alpha_n)- B \exp(-\beta \pS) =&
B \exp(-\frac{\beta}{2} \pS^\alpha_n )\left( \exp(-\frac{\beta}{2} \pS^\alpha_n )- \exp(-\frac{\beta}{2} \pS )\right)   +\\
&+ B \left( \exp(-\frac{\beta}{2} \pS^\alpha_n )- \exp(-\frac{\beta}{2} \pS )\right) \exp(-\frac{\beta}{2} \pS ) 
\end{split}
\end{equation*}

\noindent
The first term:
$$\left|\Tr\left[B \exp(-\frac{\beta}{2} \pS^\alpha_n )\left( \exp(-\frac{\beta}{2} \pS^\alpha_n )- \exp(-\frac{\beta}{2} \pS )\right)\right]\right|\leq
||B \exp(-\frac{\beta}{2} \pS^\alpha_n )|| \,\Tr \left|\exp(-\frac{\beta}{2} \pS^\alpha_n )- \exp(-\frac{\beta}{2} \pS )\right|
$$
By the previous theorem, $\displaystyle \lim_{n\rightarrow\infty}\exp(-\frac{\beta}{2} \pS^\alpha_n )= \exp(-\frac{\beta}{2} \pS )$ in $L^1(\Hcal)$. Since 
we started this proof by proving uniform (in $n$) bound for $||B \exp(-\frac{\beta}{2} \pS^\alpha_n )||$ the first term clearly goes to $0$.

\noindent The second  term:
$$\Tr\left[B \left( \exp(-\frac{\beta}{2} \pS^\alpha_n )- \exp(-\frac{\beta}{2} \pS )\right) \exp(-\frac{\beta}{2} \pS ) \right]=
\Tr\left[ \left(\exp(-\frac{\beta}{2} \pS ) B\right) \left( \exp(-\frac{\beta}{2} \pS^\alpha_n )- \exp(-\frac{\beta}{2} \pS )\right)  \right]$$
Note that, by assumptions,  $B$ is closable and  
$$\exp(-\frac{\beta}{2} \pS ) B\subset \exp(-\frac{\beta}{2} \pS ) \overline{B}= (\exp(-\frac{\beta}{2} \pS )^* B^{**}=(B^* \exp(-\frac{\beta}{2} \pS ))^*$$
We have assumed that $B^*$ is $\pT$-bounded, so by Cor.~\ref{wniosek1} the operator $B^* \exp(-\frac{\beta}{2} \pS )$ is 
bounded, as well as $\exp(-\frac{\beta}{2} \pS ) B$,  
 therefore the second term goes to $0$ as well. The proof is completed.
\hspace*{\fill}$\Box$\hspace*{1em}

Notice that due to (\ref{PM-kom}):  $\displaystyle P_n\exp(-\beta \pS^\alpha_n)P_n=P_n\exp(-\beta P_n \pS P_n )P_n$ therefore
$$\Tr\left[P_n B P_n \exp(-\beta \pS^\alpha_n)\right]=\Tr\left[P_n B P_n \exp(-\beta P_n \pS P_n) \right]$$ and we obtain, 
for $B$ and $\pS$ as in Thm.~\ref{tw-granica},
\begin{equation}\label{averages-conv}
\lim_{n\rightarrow\infty}\Tr\left[P_n B P_n \exp(-\beta P_n \pS P_n) \right]=\Tr\left[B  \exp(-\beta \pS ) \right] 
\end{equation}

\section{Bose-Hubbard system and Bogolyubov inequality}
\label{sec:BHandBIneq}

So far, we didn't make any assumptions concerning the set of sites $\La$ (besides its  finitness). Now we specify $\Lambda$ to be the cubic lattice
\begin{equation}\label{def-Lambda}
\Lambda:=\{0, 1,\dots, \lrs-1\}^d\subset \Zdb^d\subset \Rdb^d\quad,\quad d\in\Ndb
\end{equation}
We will work with `momentum' variables from the first  Brillouin zone:
\begin{equation}\label{def-Lambdadual}
\widehat{\Lambda}:=\{ \kgr\in \Rdb^d: k_i=\frac{2 \pi }{\lrs} n_i,  n_i\in \{0,\dots,  (\lrs-1)\} \}
\end{equation}
Then $|\widehat{\Lambda}|=|\Lambda|=\lrs^d$. 
At some moment we will assume that our interaction is ``translationally invariant'' and to express this condition we will think of the 
lattice $\Lambda$ as of embedded into torus (or simply as of direct product cyclic group $\Zdb_\lrs^d$) and we will add elements of $\Lambda$ as elements of $\Zdb_\lrs^d$:
\label{torus-addition}
\begin{equation}\label{torus-addition}
\xgr\dodgrup\ygr:=((x_1+y_1)\hspace{-2ex}\mod\lrs, ,\dots,(x_d+y_d)\hspace{-2ex}\mod \lrs)
\end{equation}
The  corresponding subtraction  operation will be denoted by  $\xgr\odejgrup\ygr$. Note, for  $\kgr\in \widehat{\Lambda},\,\xgr,\ygr\in\Lambda$, equalities: 
\begin{equation}
 \exp( i \kgr\cdot (\xgr\odejgrup\ygr))=\exp (i \kgr\cdot (\xgr- \ygr))\quad,\qquad \sum_\kgr e^{i\kgr\cdot\xgr}=|\Lambda|\delta_{\xgr \Zerogr}
\end{equation}
For $\kgr\in\widehat{\Lambda}$ let us define:
\begin{equation}\label{def-ck}
c(\kgr):=\frac{1}{\sqrt{|\La|}} \sum_\xgr e^{i\kgr\cdot\xgr}c_\xgr
\end{equation}
Then 
$\displaystyle 
c^\dg(\kgr)=\frac{1}{\sqrt{|\La|}} \sum_\xgr e^{-i\kgr\cdot\xgr}c^\dg_\xgr\,$ and 
\be \label{cxfromck} 
c_\xgr=\frac{1}{\sqrt{|\La|}} \sum_\kgr e^{-i\kgr\cdot\xgr}c(\kgr),
\quad
c^\dg_\xgr=\frac{1}{\sqrt{|\La|}} \sum_\kgr e^{i\kgr\cdot\xgr} c^\dg(\kgr).
\ee
Moreover, the following easy equality will be used later on: 
\be
\sum_\xgr c^\dg_\xgr c_\xgr =\sum_\kgr c^\dg(\kgr) c(\kgr). 
\label{SumRule}
\ee
Partially motivated by the paper \cite{SSZ} (where absence of superconducting order in the fermionic Hubbard model
has been proved) we define  the operators $C$ and $A$ (as operators on $\DD$) as follows:
\begin{align}\label{def-AC}
A\equiv A(\kgr) &:= c^\dg(\kgr)=\frac{1}{\sqrt{|\La|}} \sum_\xgr e^{-i\kgr\cdot\xgr}c^\dg_\xgr\,,&
C\equiv C(\kgr)&:= \frac{1}{\sqrt{|\La|}} \sum_\xgr e^{i\kgr\cdot\xgr}n_\xgr.
\end{align}

We will need various commutators, recall that all of them have the meaning as operators on $\DD$. 
Using relations (\ref{basic-rel}) by straightforward (although a little bit lengthy) computations we obtain:
\begin{equation}\label{komAAplus}
[A^\dg, A] = 1
\end{equation}
\begin{equation}\label{KomCA}
[C,A] =\frac{1}{{|\La|}}\sum_\xgr  c^\dg_\xgr= \frac{1}{\sqrt{|\La|}}c^\dg(\0gr)
\end{equation}
\begin{equation}\label{KomCN}
[C, \widehat{N}]=[C,  \widehat{N}_2]=0,
\end{equation}
\begin{equation}\label{KomCL}
[C,L]=\frac{1}{\sqrt{|\Lambda|}}\sum_\xgr e^{i\kgr\xgr}(c_\xgr^\dg-c_\xgr)\quad,\qquad
[[C,L],C^\dg]=-\frac{1}{|\La|}\sum_\xgr(c^\dg_\xgr + c_\xgr)
=-\frac{L}{|\Lambda|}
\end{equation}

\begin{align}
[C, T] &= \frac{1}{\sqrt{|\Lambda|}}\sum_{\xgr\zgr} t_{\xgr \zgr} (e^{i \kgr\xgr}-e^{i \kgr\zgr}) c_\xgr^\dg c_\zgr\nonumber\\
\label{KomCT}
[[C,T],C^\dg] 
 &= -\frac{2}{|\La|}\sum_{\xgr,\ygr} t_{\xgr\ygr}(1-\cos(\kgr\cdot(\xgr-\ygr)))c^\dg_\xgr c_\ygr 
\end{align}

\begin{equation}\label{KomCH}
\begin{split}
[[C,H],C^\dg] &= [[C,T],C^\dg]+ \lambda [[C,L],C^\dg]=\\
&= -\frac{2}{|\La|}\sum_{\xgr,\ygr} t_{\xgr\ygr}(1-\cos(\kgr\cdot(\xgr-\ygr)))c^\dg_\xgr c_\ygr -\frac{\lambda}{|\Lambda|} L
\end{split}
\end{equation}
We are going to use inequality (\ref{finiteBogol1}) for  a subspace $V:=\DM{M}$.
Recall that  $P_M$ denotes  the orthogonal projection on $\DM{M}$ and let $q_M$ be 
the orthogonal projection on $\DD_M$; there are  obvious relations 
\begin{equation}\label{def-QM}
q_M+ P_{M-1}=P_M\quad,\quad q_M P_M=q_M=P_M q_M
\end{equation} 
It is straightforward to verify that operators $A, C$ defined in (\ref{def-AC}) satisfy assumptions of Prop.~\ref{komut-finite} with
$W:=P_{M-1}(\Hcal)$ and $V:=P_M(\Hcal)$ i.e.
\begin{equation*}
P_M C\subset C P_M\,,\,\, (C_M)^*=(C^\dg)_M=C^\dg P_M\supset P_M C^\dg
\end{equation*}
\begin{equation*}
P_M A \subset  A P_{M-1}\,\,,\,(A_M)^*=(A^\dg)_M= A^\dg P_M\supset P_{M-1} A^\dg \,\,,\, 
\end{equation*}
therefore, using this proposition and notation introduced in (\ref{defBM}) we obtain:
\begin{align*}
[C_M, A_M]=[C,A]_M \,, \quad & \quad [[C_M, H_M], (C_M)^*]=[[C,H],C^\dg]_M  \\
A_M(A_M)^*+ (A_M)^*A_M&=(A A^\dg +A^\dg A)_M- q_MA^\dg A q_M, 
\end{align*}
and 
\begin{align}
\frac{\beta}{2}  \brak (A A^\dg +A^\dg A)_M\kket_M \brak[[C,H], C^\dg]_M\kket_M &\geq |\brak[C,A]_M\kket_M|^2\label{finiteBogol2}
\end{align}

Let us now, compute more explicitely terms appearing in (\ref{finiteBogol2}).

By (\ref{KomCA}):  
\begin{equation}\label{KomCMAM}
|\brak[C,A]_M\kket_M|^2= \frac{1}{|\La|}|\brak (c^\dg(\0gr))_M \kket_M|^2=:\mgr_M.
\end{equation}
The quantity $\mgr_M$ is the `` finite dimensional approximation'' of order parameter and its estimate is main goal of further considerations.

Since $0\leq \brak [[C_M, H_M], (C_M)^*]\kket_M=  \brak [[C, H], C^\dg]_M\kket_M$ we can write using (\ref{KomCH})
\begin{equation}\label{est-com3}
\begin{split}
\brak [[C, H], C^\dg]_M\kket_M &=|\brak [[C, H], C^\dg]_M\kket_M|=\\
&= \left|-\frac{2}{|\La|}\sum_{\xgr,\ygr} t_{\xgr\ygr}(1-\cos(\kgr\cdot(\xgr-\ygr)))\brak (c^\dg_\xgr c_\ygr)_M\kket_M -
\frac{\lambda}{|\Lambda|} \brak L_M \kket_M\right|\leq \\
&\leq \frac{2}{|\La|}\left|\sum_{\xgr,\ygr} t_{\xgr\ygr} (1-\cos(\kgr\cdot(\xgr-\ygr))) \brak (c^\dg_\xgr c_\ygr)_M\kket_M\right| 
+ \frac{|\lambda|}{|\Lambda|} |\brak L_M \kket_M | 
\end{split}\end{equation}
The second term in above inequality is estimated in the following
\begin{lemma}
$$| \brak L_M\kket_M |\leq \brak N_M\kket_M+|\Lambda|$$
\end{lemma}
{\em Proof:} The inequality follows from  definitions of $\displaystyle \widehat{N}=\sum_\xgr c^\dg_\xgr c_\xgr$, 
$\displaystyle L=\sum_\xgr (c^\dg_\xgr+ c_\xgr)$  
and two obvious inequalities (c.f (\ref{c-oszac}) ) : $$\sum_\xgr P_M(c_\xgr\pm 1)^\dg (c_\xgr\pm 1)P_M\geq 0$$
\hspace*{\fill}$\Box$\hspace*{1em}

To estimate the first term of the  inequality  (\ref{est-com3}) we  impose  {\em translational invariance} condition. To this end, let us start by 
``changing  variables'' of summation: 
$$\Lambda\times\Lambda\ni (\xgr,\ygr)\mapsto (\xgr, \ygr \odejgrup \xgr)=:(\xgr, \zgr) \in\Lambda\times\Lambda.$$
Since $\cos \kgr\cdot (\xgr-\ygr)=\cos \kgr \cdot (\xgr\odejgrup\ygr)$ we have 
$$\sum_{\xgr,\ygr} t_{\xgr\ygr} \,(1-\cos(\kgr\cdot(\xgr-\ygr)))\, c^\dg_\xgr c_\ygr =
\sum_{\xgr,\zgr}t_{\xgr,\xgr\dodgrup\zgr}\,(1-\cos(\kgr\cdot\zgr))\, c^\dg_\xgr c_{\xgr\dodgrup\zgr}$$
Our  translational invariance condition means:
\begin{equation}\label{translational}
\forall\, \xgr, \ygr,\zgr \in\Lambda \,:\quad  t_{\xgr,\xgr\dodgrup\zgr}=t_{\ygr,\ygr\dodgrup\zgr}=:\tilde{t}_\zgr
\end{equation}
Let $\MM_2$  satisfies:  
\begin{equation}\label{def-k2}
\sum_{\zgr}| \tilde{t}_\zgr||\zgr|^2\leq \MM_2
\end{equation}
Later on, in thermodynamic limit, 
we will assume that $\MM_2$ can be chosen independently of $\Lambda$ (cf. Remark.~\ref{stale-M}).
Using translational invariance  together with (\ref{cxfromck}) we compute 
\begin{equation*}
\begin{split}
\sum_{\xgr,\zgr} & \tilde{t}_\zgr (1-\cos(\kgr\cdot\zgr)) c^\dg_\xgr c_{\xgr\dodgrup\zgr}= 
\sum_{\zgr}\tilde{t}_\zgr (1-\cos(\kgr\cdot\zgr)) 
\sum_{\xgr}\frac{1}{|\Lambda|}\sum_{\jgr}\sum_{\lgr}\ e^{i(\jgr-\lgr)\cdot \xgr} e^{-i\lgr\cdot \zgr}c^\dg(\jgr)c(\lgr)=\\
&=\sum_{\zgr,\lgr}\tilde{t}_\zgr (1-\cos(\kgr\cdot\zgr)) e^{-i\lgr\cdot \zgr}c^\dg(\lgr)c(\lgr)
\end{split}
\end{equation*}
and the first term of the  inequality  (\ref{est-com3}) can be estimated as
\begin{equation*}
\begin{split}
&\left|\sum_{\xgr,\ygr} t_{\xgr\ygr} (1-\cos(\kgr\cdot(\xgr-\ygr))) \brak (c^\dg_\xgr c_\ygr)_M\kket_M\right| =
\left|\sum_{\zgr,\lgr}\tilde{t}_\zgr (1-\cos(\kgr\cdot\zgr)) e^{-i\lgr\cdot \zgr} \brak (c^\dg(\lgr)c(\lgr))_M\kket_M\right|\leq\\
&\leq \sum_{\zgr,\lgr}| \tilde{t}_\zgr|\, (1-\cos(\kgr\cdot\zgr)) | \brak (c^\dg(\lgr)c(\lgr))_M\kket_M|
\leq \frac12 |\kgr|^2 | \brak \widehat{N}_M \kket_M |\,\sum_{\zgr}| \tilde{t}_\zgr||\zgr|^2\leq \frac12 \MM_2 |\kgr|^2  \brak \widehat{N}_M \kket_M,
\end{split}
\end{equation*}
where elementary inequalities $1-\cos x \leq\frac{1}{2} x^2$,  $(\kgr\cdot\zgr)^2\leq |\kgr|^2 |\zgr|^2$  together with  the 
equality  $| \brak (c^\dg(\lgr)c(\lgr))_M\kket_M|= \brak (c^\dg(\lgr)c(\lgr))_M\kket_M$, and (\ref{SumRule}) were used.

This way we obtain the estimate:
$$\brak [[C, H], C^\dg]_M\kket_M \leq  |\lambda|+ \frac{\brak \widehat{N}_M \kket_M}{|\Lambda|} (\MM_2 |\kgr|^2+ |\lambda|).$$ 
Notice that by (\ref{komAAplus}): 
\begin{equation}\label{AAplusanty}
\brak (A A^\dg +A^\dg A)_M\kket_M = \brak ( 2 A A^\dg +1)_M\kket_M=\brak (  2 c^\dg(\kgr) c(\kgr))_M\kket_M +1 
\end{equation}
Now, by (\ref{KomCMAM}, \ref{AAplusanty}) we can write the Bogolyubov inequality (\ref{finiteBogol2}) as: 
\begin{equation*}
\mgr_M
\leq \frac{\beta}{2}\left(\brak (  2 c^\dg(\kgr) c(\kgr))_M\kket_M +1 \right)
\left(|\lambda|+ \frac{\brak \widehat{N}_M \kket_M}{|\Lambda|} (\MM_2 |\kgr|^2+ |\lambda|)\right)
\end{equation*}

Let us introduce a ``finite dimensional approximation'' of density 
\begin{equation}\label{def-roM}
\rho_M:=\frac{\brak \widehat{N}_M \kket_M}{|\Lambda|}.
\end{equation}
Now the Begolyubov inequality reads:
$$\frac{\mgr_M}{|\lambda|+ \rho_M (\MM_2 |\kgr|^2+ |\lambda|)}
\leq \frac{\beta}{2}\left(\brak (  2 c^\dg(\kgr) c(\kgr))_M\kket_M +1 \right).$$
Summing over $\kgr\in \hat{\Lambda}$ using (\ref{SumRule}) and dividing by $|\Lambda|$ we get:
\begin{equation}\label{Bogol2}
\mgr_M \frac{1}{|\Lambda|}\sum_{\kgr}\frac{1}{\rho_M \MM_2 |\kgr|^2+ |\lambda|(\rho_M+1)}\leq \beta( \rho_M+\frac12)
\end{equation}
Now we shall show that $\mgr_M$ and $\rho_M$ have limits as $M\rightarrow\infty$:

\begin{proposition}\label{prop-limit}
\begin{equation*}
\lim_{M\rightarrow \infty}\brak \widehat{N}_M \kket_M = \brak \widehat{\NN} \kket\quad,\qquad 
\lim_{M\rightarrow \infty}\brak (c^\dg(0))_M \kket_M = \brak (c(0))^* \kket
\end{equation*}
\end{proposition}
{\em Proof:} We are going to use Thm.~\ref{tw-granica} with the following identifications:
$\pT:=\UU\widehat{\NN}_2$ (then  $\pT|_\DD=\UU\widehat{N}_2$),  $A:=T-\mu\widehat{N}+\lambda L$  
and our sequence of projections is $P_M$.  By Prop.~\ref{prop-N2-estimates} the operator $A$ is $\UU\widehat{N}_2$-bounded with relative 
bound $0$, so clearly we can choose $b<\frac13$ and we have a self-adjoint operator $\pH(u)$.

We know that $\widehat{N}$ is $\UU\widehat{N}_2$-bounded, so (compare the proof of Prop.~\ref{TW-pomoc}) 
$\widehat{\NN}=\widehat{\NN}^*$ is $\UU\widehat{\NN}_2$-bounded, so it can be put as operator $B$ in  Thm.~\ref{tw-granica}.

By the Lemma~\ref{c-cplus-est} $c(0)$ and $c^\dg(0)$ are $\UU\widehat{N}_2$ bounded. 
Clearly $c(0)=(c^\dg(0))^*|_\DD$ and $c^\dg(0)=(c(0))^* |_\DD$ so both of them are closable and their closures are $\UU\widehat{\NN}_2$-bounded, 
so each  of them can be used as $B$ in Thm.~\ref{tw-granica}.
So, as in (\ref{averages-conv}):
$$\lim_{M\rightarrow\infty}\Tr\left[P_M \widehat{N} P_M \exp(-\beta P_M \pH(\UU) P_M) \right]=\Tr\left[\widehat{\NN}  \exp(-\beta \pH(\UU) ) \right]$$ 
$$\lim_{M\rightarrow\infty}\Tr\left[P_M c^\dg(0) P_M \exp(-\beta P_M \pH(\UU) P_M) \right]=\Tr\left[(c(0))^*  \exp(-\beta \pH(\UU) ) \right]$$ 
$$\lim_{M\rightarrow\infty}\Tr\left[P_M \exp(-\beta P_M \pH(\UU) P_M) \right]=\Tr\left[\exp(-\beta \pH(\UU) ) \right].$$ 
The proposition follows.
\hspace*{\fill}$\Box$\hspace*{1em}

Let us denote (c.f. (\ref{KomCMAM},\ref{def-roM}))
$$\mgr_\Lambda:= \lim_{M\rightarrow\infty} \mgr_M=\frac{|\brak (c(0))^* \kket|^2}{|\Lambda|}\quad{\rm  and}\quad  
\rho_\Lambda:=\lim_{M\rightarrow\infty} \rho_M=\frac{\brak \widehat{\NN} \kket}{|\Lambda|}.$$
Clearly $\mgr_\Lambda\geq 0 $ and $\rho_\Lambda >0$, note that both quantities depend on $\lambda$ 
(and other parameters as well, but they are considered fixed) and we will write $\mgr_\Lambda(\lambda), \rho_\Lambda(\lambda)$ 
when we want to stress this dependence.
Now, passing to the limit $M\rightarrow\infty$ in (\ref{Bogol2}) and multiplying by  $\rho_\Lambda$ we  obtain:
\begin{equation}\label{almost-final-ineq}
\mgr_\Lambda \frac{1}{|\Lambda|} \sum_{\kgr}\frac{1}{  \MM_2 |\kgr|^2+ |\lambda|( 1 +1/\rho_\Lambda )}\leq \beta \rho_\Lambda  \left( \rho_\Lambda+\frac12\right)
\end{equation}

\section{Thermodynamic limit}
\label{sec:TDLim}

At this moment it is useful to state clearly assuptions we made before passing to the limit. 
We will consider only Hamitonians defined by (\ref{HamWZK}), i.e.
\be
H_\La(\mu, \lambda) := \UU \,\widehat{N}_{2,\La}  +T_\La  - \mu\, \widehat{N}_\La + \lambda L_\La  
\ee
In the thermodynamic limit dependence on $\mu$ and $\lambda$ is important, so we underline it in notation.
We assume the following conditions are satisfied:
\begin{enumerate}
\item The lattice $\Lambda$ is a cubic lattice as in (\ref{def-Lambda}); 
\item Translational invariance of $T$  as  in (\ref{translational}): $\displaystyle t_{\xgr,\xgr\dodgrup\zgr}=:\tilde{t}_\zgr$;
\item There exists $\MM_2$ {\em independent } of $\Lambda$  such that $\displaystyle \sum_{\zgr\in\Lambda}| \tilde{t}_\zgr||\zgr|^2\leq \MM_2$;
\item There exists $\MM_d$  as in (\ref{def-M}) {\em independent } of $\Lambda$ and  $|\tilde{t}_0|\leq \MM_d$. 
\end{enumerate}
\begin{remark}Conditions {\rm (2-4)} imply that $\MM$ defined in {\rm (\ref{def-M})}  can be chosen independently of $\Lambda$;
\end{remark}

\noindent
Te get our estimates for $\mgr_\Lambda$ we need uniform in $\lambda$ and $\Lambda$ bounds on density $\rho_\Lambda$: 
\begin{proposition}\label{prop-density} For every $\mu\in\Rdb$, there exist strictly positive numbers $\lambda_0,\,\rho_1,\,\rho_2$ 
such that for every lattice $\Lambda$:
$$\left(\,\, |\lambda|\leq \lambda_0\,\,\right)\,\Rightarrow \,\left(\,\,\rho_1\leq \rho_\Lambda(\lambda)\leq \rho_2\,\,\right)$$
\end{proposition}
This  proposition is proven at the end of this section. 

\noindent
Now, by (\ref{almost-final-ineq}) and equality $|\Lambda|=\lrs^d$ we have the estimate  
\begin{equation}\label{final-ieq}
\begin{split}
\mgr_\Lambda \left(\frac{2\pi}{\lrs}\right)^d\sum_{\kgr}\frac{1}{\MM_2 |\kgr|^2+ |\lambda|(1+1/\rho_1)}
&\leq \mgr_\Lambda \left(\frac{2\pi}{\lrs}\right)^d\sum_{\kgr}\frac{1}{\MM_2 |\kgr|^2+ |\lambda|(1+1/\rho_\Lambda)}\leq\\
&\leq (2\pi)^d \beta \rho_\Lambda \left( \rho_\Lambda+\frac12\right) \leq (2\pi)^d \beta \rho_2 \left( \rho_2+\frac12\right)
\end{split}
\end{equation}

\noindent
It is easy to see, that when  $\lrs\rightarrow\infty$ the quantity 
$\displaystyle \left(\frac{2\pi}{\lrs}\right)^d\sum_{\kgr}\frac{1}{\MM_2 |\kgr|^2+ |\lambda|(1+1/\rho_1)}$ converges to the integral
$$\int_{[0,2\pi]^d} \frac{ d^d \kgr}{ \MM_2 |\kgr|^2+ \alpha }\,,\,\,\,\alpha:=|\lambda|(1+1/\rho_1)$$
This integral for $d\geq 3$ is finite for any $\lambda\in\Rdb$ and 
\begin{align*}
&{\rm for}\quad d=1 : \qquad  \int_{[0,2\pi]} \frac{ d x}{ \MM_2 x^2 + \alpha}=
\frac{1}{\sqrt{\MM_2\alpha}}\,\arctan \left(2\pi\sqrt{\frac{\MM_2}{\alpha}}\right)\,,\\
&{\rm for}\quad d=2: \qquad \int_{[0,2\pi]^2} \frac{ d x d y }{ \MM_2  (x^2+y^2) + \alpha }\geq \int_0^{\pi/2} d\varphi \int_0^{2\pi}
\frac{ r dr  }{ \MM_2  r^2 + \alpha}=\frac{\pi}{4 \MM_2}\log\left(1+\frac{4 \MM_2 \pi^2}{\alpha}\right).
\end{align*}
Clearly  both integrals are divergent  as  $\lambda \rightarrow 0$  and  finally we  get ``no condensation'' result:
\begin{proposition}
Let $d=1$ or $d=2$ and $N$ be the linear size of the (cubic) lattice  $\Lambda$, i.e. $|\Lambda|=N^d$. 
For any $\epsilon>0$ there exist $\delta_\epsilon >0 $ and $\lrs_\epsilon$ such that: 
$$\left(\,\,|\lambda| <\delta_\epsilon \,\, {\rm and} \,\,  \lrs \geq \lrs_\epsilon\,\,\right)\, \Rightarrow\, |\mgr_\Lambda(\lambda)|\leq \epsilon$$ 
\end{proposition}
\hspace*{\fill}$\Box$\hspace*{1em}

\subsection{Proof of proposition \ref{prop-density}}

In the following we will use  Thm. \ref{tw-granica} and formula (\ref{averages-conv}) for the hamiltonian
\begin{equation}
\pH_\Lambda(\mu,\lambda):=\UU \,\widehat{\NN}_{2,\La}  +\overline{T_\La  - \mu\, \widehat{N}_\La + \lambda L_\La }
\end{equation}
The density $\rho_\Lambda(\mu,\lambda)$ is given by:
\begin{equation}
\rho_\Lambda(\mu,\lambda) =\frac{1}{|\Lambda|}\frac{\Tr[\widehat{\NN}_\Lambda \exp(-\beta \pH_\Lambda(\mu,\lambda))]}{ \Tr[\exp(-\beta \pH_\Lambda(\mu,\lambda))]}
\end{equation}
To simplify notation we will write $\pH(\mu)$ keeping in mind dependence on other parameters.

We shall follow ideas of \cite{Ginibre} and \cite{Ruelle-Helvetica} 
and begin by proving that the density can be expressed as derivative with respect to $\mu$ of 
$\log \Tr[\exp(-\beta \pH(\mu)]$, and then use convexity arguments. The following lemma collects useful  known results. 
 

\begin{lemma}\label{lema-Ginibre}
Let $V$ be a {\em finite dimensional} Hilbert space. \\
1) For any  $F,G\in \Bcal(V)\,$:  $\displaystyle \frac{d}{ d t}\Tr(\exp(F + t G))=\Tr(G\exp(F + t G))$\\
2) For any {\em self-adjoint} $F,G\in \Bcal(V)\,$:  $|\Tr(\exp(F + i G))|\leq \Tr(\exp(F))$\\
3) The function $A\mapsto \log \Tr (\exp(A))$ is convex and increasing on the real space of self-adjoint elements of $\Bcal(V)$.
\end{lemma}
{\em Proof:} The first statement is clear; the second one is Lemma 2 from \cite{Ginibre}. 
 the third one is
Prop. 2.5.5 in \cite{Ruelle}
\\
\hspace*{\fill}$\Box$\hspace*{1em}

\noindent
We will also need the following result: 


\begin{lemma}\label{lemat-Ruelle}\cite{Ruelle-Helvetica}
Let $A$ be bounded from below and  self-adjoint operator on $\Hcal$. Assume moreover that $\exp(-A)$ is trace class. Let 
$\varphi_j\,,\,j=1,\dots, n$ be an orthonormal sequence in $D_A$. Then
$$\sum_{i=1}^n \exp(-(\varphi_i\,|\, A \varphi_i))< \Tr(\exp(-A))$$
\end{lemma}
\hspace*{\fill}$\Box$\hspace*{1em}

\noindent
Let $P_M$ be the family of projections on  $\DM{M}$ (c.f. (\ref{def-DM})) and  let us define 
$$H_M(\mu):=P_M\pH(\mu) P_M=P_M(\UU \,\widehat{N}_2  +T + \lambda L)P_M   - \mu\, P_M \widehat{N} P_M=: \widehat{G}_M -\mu \widehat{N}_M, $$
and  the family of entire functions:
$$\Cdb\ni \mu\mapsto f_M(\mu):=\Tr\left[P_M\exp(- \beta H_M(\mu))P_M\right]=\Tr\left[P_M\exp(- \beta \widehat{G}_M+ \beta\mu \widehat{N}_M)P_M\right]$$
Clearly for {\em real} $\mu$ these functions are strictly positive  and, by (\ref{averages-conv}), 
$\displaystyle \lim_{M\rightarrow \infty} f_M(\mu)=\Tr(\exp[-\beta \pH(\mu)])$.
We are going to use Vitali-Porter Theorem (\cite{Schiff}, p. 44)
to prove that this convergence is almost-uniform,  therefore the function
\begin{equation} \label{def-fun-f} 
\Rdb\ni\mu\mapsto f(\mu):=\Tr(\exp(-\beta \pH(\mu)))
\end{equation} 
would extend to an entire function. 
Then the sequence of derivatives $f'_M$ would converge almost-uniformly to $f'$. On the other hand, by Lemma \ref{lema-Ginibre},  
$f'_M(\mu)=\beta \Tr(\widehat{N}_M \exp[-\beta H_M(\mu)])$ and  this sequence  converges, for {\em real} $\mu$,  
to $\beta \Tr(\widehat{\NN} \exp[-\beta \pH(\mu)])$ by (\ref{averages-conv}). 

\noindent To use the Vitali's Theorem we will show
\begin{lemma}
Let $Re(\mu)\leq c$. There exists $C>0$ such that for any $M\in\Ndb:\,  |f_M(\mu)|\leq C$.
\end{lemma}
{\em Proof:} By (2) of Lemma \ref{lema-Ginibre}:   $|f_M(\mu)|\leq f_M(Re(\mu))$. 
Since  $f_M'(\mu)=\beta \Tr(\widehat{N}_M \exp[-\beta H_M(\mu)])$, and this quantity is 
positive  for {\em real} $\mu$,  the function $f_M(\mu)$ is increasing on the real axis therefore $|f_M(\mu)|\leq f_M(c)$. We are going to prove that
for a given $\mu\in\Rdb$ the sequence $(f_M(\mu))$ is increasing. 

To this end, let  $\varphi_1,\dots,\varphi_l\in P_M\Hcal\subset P_{M+1}\Hcal$ be an orthonormal basis of eigenvectors of $H_M(\mu)\,,\,\mu\in\Rdb$ 
with eigenvalues $\lambda_k$:
$H_M(\mu) \varphi_k=\lambda_k\varphi_k$. Then
\begin{equation*}
\begin{split}
f_M(\mu)=&\sum_{k=1}^l (\varphi_k\,|\,e^{-\beta \lambda_k} \varphi_k)=\sum_{k=1}^l e^{-\beta \lambda_k}=
\sum_{k=1}^l \exp[(\varphi_k\,|\, -\beta H_M \varphi_k)]=\\=&\sum_{k=1}^l \exp[(\varphi_k\,|\, -\beta H_{M+1} \varphi_k)]< 
\Tr(\exp(-\beta H_{M+1}))=f_{M+1}(\mu)
\end{split}
\end{equation*}
where the last inequality follows from  the  lemma \ref{lemat-Ruelle}.\hspace*{\fill}$\Box$\hspace*{1em}\\
Then, since 
$\displaystyle \lim_{M\rightarrow\infty}f_M(c)=\Tr(\exp[-\beta \pH(c)])$,  we may put $C:=\Tr(\exp[-\beta \pH(c)])$.

\noindent
This way we have proven that for $\mu\in\Rdb$:
$\displaystyle \frac{d}{d \mu}  \Tr(\exp[-\beta \pH(\mu)])= \beta \Tr(\widehat{\NN} \exp[-\beta \pH(\mu)])$ and
$$\rho_\Lambda(\mu)=\frac{1}{\beta|\Lambda|}\frac{d}{d \mu} \log f(\mu)= \frac{1}{\beta|\Lambda|}\frac{d}{d \mu}  \log \Tr[\exp[-\beta \pH(\mu)])$$

Now we are going to use convexity arguments to get estimates for $\rho_\Lambda(\mu)$ . 
From (3) of Lemma \ref{lema-Ginibre} it follows that each of functions 
$\displaystyle \Rdb\ni\mu\mapsto \log f_M(\mu)$ is convex.  Therefore the function 
\begin{equation}\label{def-R}
R_{\Lambda,\lambda}:
\Rdb\ni\mu\mapsto \frac{1}{\beta|\Lambda|} \log f(\mu)= \frac{1}{\beta|\Lambda|}\log \Tr(\exp[-\beta \pH(\mu)])
\end{equation}
is convex  and increasing (because  $f$ is increasing). We will need the lemma which is (simplified for our needs version of) Lemma 3 in \cite{Ruelle-Helvetica}:
%
%
\begin{lemma}\label{Ruelle-Helvetica} Let $\DD$ be dense in $\Hcal$ and $A,B:\DD\rightarrow \Hcal$ be essentially self-adjoint. 
Assume that $B$ be bounded from below,  $\exp(-B)$ is  trace-class and that there  exists $\alpha\geq 0$ such that for every $\psi\in\DD$:
$\displaystyle\,(\psi \,|\, (A-B)\psi)\geq\alpha \,(\psi\,|\, \psi).\,$
Then $$\Tr(\exp(-A))\leq e^{-\alpha}\,\Tr(\exp(-B)).$$
\hspace*{\fill}$\Box$\hspace*{1em}
\end{lemma} 

\noindent
To apply this lemma let us observe that by (\ref{c-oszac}) and (\ref{c-T-oszac}) we have
\begin{lemma} Let $\DD$ be as defined in {\rm (\ref{def-D})}  and  $\MM$ be as in {\rm (\ref{def-M})}.  Then for every $\psi\in\DD$:
$$-\MM \left(\psi \,|\, \hat{N} \psi\right) \leq  
\left(\psi \,|\, T \psi\right) \leq \MM \left(\psi \,|\, \hat{N} \psi\right)$$
$$-|\lambda| \left(\psi \,|\, (\hat{N}+|\Lambda|) \psi\right) \leq  
\left(\psi \,|\, \lambda L \psi\right) \leq |\lambda|  \left(\psi \,|\, (\hat{N}+|\Lambda|) \psi\right).$$
\end{lemma}
\hspace*{\fill}$\Box$\hspace*{1em}

\noindent
For $\psi\in\DD$:
\begin{equation*}
\begin{split}
(\psi\,|\, [\UU \,\widehat{N}_{2} - & (\mu+\MM +|\lambda|)\,\widehat{N}] \psi) - |\lambda| |\Lambda| \, (\psi\,|\,\psi) \leq \\
\leq &(\psi\,|\,\pH(\mu) \psi) \leq 
 (\psi\,|\, [\UU \,\widehat{N}_{2} - (\mu-\MM -|\lambda|)\,\widehat{N}] \psi)  + |\lambda| |\Lambda| \, (\psi\,|\,\psi)
\end{split}
\end{equation*}
For $r\in\Rdb$ let us define the  operator $\widehat{G}(r): \DD\rightarrow \Hcal $ by 
$$\widehat{G}(r):=\UU \,\widehat{N}_{2} -r  \,\widehat{N} =\sum_{\xgr \in\Lambda}\UU \,\hat{n}^2_{\xgr}- r\, \hat{n}_{\xgr}$$ 
and by $\widehat{\pG}(r)$ its closure.  
Remembering that $[\hat{n}_{\xgr}\,,\,\hat{n}_{\ygr}]=0$ we have
$$\dsp \exp(-\beta\widehat{\pG}(r)=\prod_{\xgr\in\Lambda}\exp[-\beta( \UU \,\hat{n}^2_{\xgr}- r\, \hat{n}_{\xgr})].$$
The operator $\dsp \exp(-\beta\widehat{\pG}(r))$ is trace-class and using the basis (\ref{baza}) 
consisting of eigenvectors of  $\widehat{\pG}(r)$ its trace can be explicitely computed:
\begin{equation}\label{trace-G}\Tr\exp(-\beta\widehat{\pG}(r))=\left(\sum_{n=0}^\infty e^{-\beta (\UU n^2 - r n)}\right)^{|\Lambda|}> 1
\end{equation}

Now, by the lemma \ref{Ruelle-Helvetica} we have inequalities:
$$\exp( -\beta|\lambda| |\Lambda|)\, \Tr\exp(-\beta\widehat{\pG}(r_1))\leq \Tr\exp(-\beta\pH(\mu))\leq 
\exp( \beta|\lambda| |\Lambda|)\, \Tr\exp(-\beta\widehat{\pG}(r_2)), $$ 
where we use abbreviations:
\begin{equation}\label{def-r1r2}
r_1:=\mu-\MM-|\lambda| \quad , \quad r_2:=\mu+\MM+|\lambda|.
\end{equation} 
Using (\ref{trace-G})
we can write 
$$ -\beta|\lambda| |\Lambda| + |\Lambda| \log \left(\sum_{n=0}^\infty e^{-\beta (\UU n^2 - r_1 n)}\right)\leq \log \Tr\exp(-\beta\pH(\mu))\leq 
 +  \beta|\lambda| |\Lambda| + |\Lambda|  \log \left(\sum_{n=0}^\infty e^{-\beta (\UU n^2 - r_2 n)}\right),
$$
and denoting
\begin{equation}\label{def-g}
g(r):=\frac1{\beta}\log \left(\sum_{n=0}^\infty e^{-\beta (\UU n^2 -r n)}\right)> 0
\end{equation}
%
we obtain for any $\mu\in\Rdb$  (c.f. (\ref{def-R}))
\begin{equation}\label{g-R-estimate}
- |\lambda| + g(r_1)
\leq R_{\Lambda,\lambda }(\mu)
\leq
|\lambda|+ g(r_2).
\end{equation}
This inequality will be used to get bounds for  $\rho_\Lambda(\lambda, \mu)=\frac{d}{d \mu}  R_{\Lambda,\lambda}(\mu)$. 

\begin{remark} One can easily verify that 
the function   $g: \Rdb\rightarrow\Rdb$ is  strictly increasing and   strictly convex. 
\end{remark}

\begin{lemma}\label{lemma-fg}
Let $f, g:\Rdb\rightarrow \Rdb$. Assume that $f$ is differentiable, $f'$ is increasing  and  $f\leq g$ 
(the prime denotes the derivative).
Then, for any $\mu_0\in \Rdb $:
$$\sup\left\{\frac{y- g(\mu)}{\mu_0-\mu }: y\leq f(\mu_0), \mu< \mu_0 \right\} \leq f'(\mu_0)\leq \inf\left\{\frac{g(\mu) -y}{\mu-\mu_0}: y\leq f(\mu_0), \mu>\mu_0 \right\}$$
\end{lemma}

\noindent
{\em Proof:}  Let $\mu_0\in \Rdb$ and $y\leq f(\mu_0)$. Then  for  $\mu<\mu_0$:
$$y-g(\mu)\leq f(\mu_0)- g(\mu)\leq f(\mu_0)- f(\mu)= f'(\xi) (\mu_0-\mu) \leq f'(\mu_0) (\mu_0-\mu);$$
and for $\mu> \mu_0$:
$$g(\mu)-y\geq  f(\mu)-f(\mu_0)= f'(\xi)(\mu-\mu_0)\geq f'(\mu_0)(\mu-\mu_0)$$
\hspace*{\fill}$\Box$\hspace*{1em}


\noindent The next lemma gives estimates we need:
\begin{lemma}\label{last-lemma}
Let $F:\, ]0, \infty[\times \Rdb\ni(\lambda,\mu)\mapsto F(\lambda,\mu)\in\Rdb$ and $ G:\Rdb\rightarrow\Rdb$. 
Denote by $F_\lambda$ the function $\Rdb\ni\mu\mapsto F(\lambda,\mu)$. Assume that:\\
1) $\forall\, \lambda >0 \,: \,F_\lambda',\, G'>0$
 (where $F_\lambda'(\mu)\equiv \frac{\df F_\lambda}{\df \mu}(\mu))$\\
2) $ \forall\, \lambda >0 \,:  F_\lambda'\,,\,G'$ are  increasing\\
3) $G$ is bounded from below;   then it is bounded by $\dsp \lim_{\mu\rightarrow -\infty} G(\mu)=:C$\\
4) There exists  $M>0$ such that for {\em every} $\lambda >0$ and every $ \mu\in \Rdb$:
$$G(\mu- M- \lambda) - \lambda\,\leq\, F_\lambda (\mu)\,\leq\, G(\mu+M+\lambda)+\lambda$$
Then, for every  $\mu_0\in\Rdb$ there exist $\lambda_0>0$ and $0< \rho_1<\rho_2<\infty$ such that
$$\forall\,\, \lambda\in]0,\lambda_0[  \quad   \rho_1 \leq F_\lambda'(\mu_0) \leq \rho_2 $$
\end{lemma}

\noindent
{\em Proof:}  For $\mu_0\in \Rdb$ let us define numbers $P_1,P_2$ and functions $Q_1, Q_2$:
$$P_1:=G(\mu_0-M)\,,\, P_2:=G(\mu_0+M)\,,\quad
Q_1(\lambda):=G(\mu_0-M-\lambda)-\lambda\,, Q_2(\lambda):=G(\mu_0+M+\lambda)+\lambda.$$
Then we have 
$$  Q_1(\lambda)< P_1 < P_2< Q_2(\lambda)\,,\quad Q_1(\lambda) \leq F_\lambda(\mu_0)\leq Q_2(\lambda)\,,\quad P_1>C.$$
It is straightforward to check that for any $\mu\in \Rdb$ the function 
$\dsp ]0, \infty[\ni\lambda\mapsto G(\mu-M-\lambda)-\lambda$ is decreasing and the function 
$\dsp ]0, \infty[\ni\lambda\mapsto G(\mu+M+\lambda)+\lambda$ is increasing therefore $Q_1(\lambda)$ is decreasing and $Q_2(\lambda)$ is increasing.
 

\noindent
Let us now prove:
$$ \left( \,\,\lambda\in \left]0, \frac{P_1-C }{2+G'(\mu_0-M)}\right[\,\,\right)\,\Rightarrow \,\left(\,\,Q_1(\lambda) > C+\lambda\,\,\right)$$
Indeed, since  $G'$ is increasing we can write 
$$Q_1(\lambda)-P_1=G(\mu_0-M-\lambda)-\lambda - G(\mu_0-M)=  -G'(\mu_0-M-\xi\lambda) \lambda -\lambda \geq  -G'(\mu_0-M) \lambda -\lambda$$
for some $\xi\in[0,1]$. 
This  way:%
$$Q_1(\lambda) \geq  P_1 - \lambda\,[G'(\mu_0 -M)+1] = P_1- \lambda\,[G'(\mu_0 -M)+2])+ \lambda >C+\lambda $$

Now, let us fix some $\lambda_0$ satysfying inequality  $\dsp 0< \lambda_0 < \frac{P_1-C }{2+G'(\mu_0-M)} $. 
Then for {\em every}  $\lambda\in]0,\lambda_0[$ and {\em every $\mu$}  we have
$$F_\lambda(\mu)\leq G(\mu+M+\lambda) +\lambda < G(\mu+M+\lambda_0) +\lambda_0=:G_0(\mu)$$ and
$$C+\lambda_0< Q_1(\lambda_0)< Q_1(\lambda)\leq F_\lambda(\mu_0)$$
This way   we have  the inequality $\dsp F_\lambda  < G_0\,$ 
for any $\lambda<\lambda_0$:

\noindent
Because $\dsp \lim_{\mu\rightarrow -\infty}G_0(\mu)=C+ \lambda_0 < Q_1(\lambda_0)\,,\,$  there is  $\tilde{\mu}<\mu_0$ such that 
$G_0(\tilde{\mu}) < Q_1(\lambda_0)$ and   $\dsp \rho_1: =\frac{Q_1(\lambda_0)-G_0(\tilde{\mu})}{\mu_0 - \tilde{\mu}} $ is strictly positive. Clearly
$\dsp \rho_1\in \left\{\frac{y- G_0(\mu)}{\mu_0-\mu }: y\leq F_\lambda(\mu_0), \mu< \mu_0 \right\}.$
Applying   lemma \ref{lemma-fg} for $f=F_\lambda$ and $g:=G_0$, we get $F_\lambda'(\mu_0) \geq\rho_1$ for every $\lambda < \lambda_0$. 

Let  $\rho_2:=G_0(\mu_0+1)-Q_1(\lambda_0)$, since $Q_1(\lambda_0)< F_\lambda(\mu_0)$, 
$\dsp \rho_2\in \left\{\frac{G_0(\mu) -y}{\mu-\mu_0}: y\leq F_\lambda(\mu_0), \mu>\mu_0 \right\}$. 
Therefore  $F_\lambda'(\mu_0) < \rho_2$ for every $\lambda < \lambda_0$. 
The proof is done.
\hspace*{\fill}$\Box$\hspace*{1em}

Finally, the   proposition \ref{prop-density} follows from the lemma \ref{last-lemma} applied to functions
$F(\lambda,\mu):=R_{\Lambda,\lambda}(\mu)$ and $G(\mu):=g(\mu)$, where $R_{\Lambda,\lambda}(\mu)$ is defined in (\ref{def-R}) and  $g$  in (\ref{def-g}).
\section{Summary}
\label{sec:Summary}
We have rigorously shown that for one- and two-dimensional Bose--Hubbard model, with translationally invariant interaction and periodic boundary conditions, 
with arbitrary hoppings, which fall-off suffciently fast with  a distance, for any filling and any positive temperature,
there  is no Bose--Einstein condensation. 
We allowed arbitrary occupation of every site. For models where the occupation is bounded by some constant,
we have also absence of Bose--Einstein condensation, as it is rather easy corollary from considerations in the paper (\cite{BFU}.

In the area of bosonic Hubbard model, 
there is  still quite a few problems, waiting for rigorous treatment (the situation is similar for fermionic version). 
Even refraining ourselves to two dimensions and positive temperature, one encounters open problems; one of the most important ones is proving 
existence of the {\em the Kosterlitz--Thouless type transition} \cite{KT}. It has been rigorously established for some  classical models 
(XY, sine-Gordon, Coulomb gas ones) \cite{FrohlichSpencer}. Despite extensive numeric and non-rigorous treatment, 
the proofs for quantum models, including 2d Bose-Hubbard one,  are  -- as far as we know -- still lacking. 

\end{document}